\documentclass[journal]{IEEEtran}
\usepackage{mathtools}
\usepackage{amsmath}
\usepackage{amsthm}
\usepackage{amssymb}
\usepackage{mathrsfs}
\usepackage{graphicx}
\usepackage{subfigure}
\usepackage{bm}
\usepackage{color}
\usepackage{cite}

\usepackage{tikz}

\newtheorem{asmp}{\textbf{Assumption}}
\newtheorem{lem}{\textbf{Lemma}}
\newtheorem{prop}{\textbf{Proposition}}
\newtheorem{thm}{\textbf{Theorem}}
\newtheorem{rmk}{\textbf{Remark}}
\newtheorem{defn}{\textbf{Definition}}

\newtheorem*{prof}{\emph{\textbf{Proof}}}

\allowdisplaybreaks[4]

\ifCLASSINFOpdf
\else
\fi
\begin{document}
%
{\title{Recursive Network Estimation\\From Binary-Valued Observation Data}}
%
%
%

\author{Yu Xing,
Xingkang He, 
Haitao Fang, 
and Karl Henrik Johansson
\thanks{This work is supported by National Key R\&D Program of China (2016YFB0901900), National Natural Science Foundation of China (61573345), Knut \& Alice Wallenberg Foundation, and Swedish Research Council.}%
\thanks{Yu Xing and Haitao Fang are with Key Laboratory of Systems and Control, Academy of Mathematics and Systems Science, Chinese Academy of Sciences, Beijing 100190, and School of Mathematical Sciences, University of Chinese Academy of Sciences, Beijing 100049, P. R. China
(e-mail: yxing@amss.ac.cn; htfang@iss.ac.cn).}
\thanks{Xingkang He and Karl Henrik Johansson are with Division of Decision and Control Systems, School of Electrical Engineering and Computer Science, KTH Royal Institute of Technology, SE-10044 Stockholm, Sweden
(e-mail: xingkang@kth.se; kallej@kth.se).}}

%
%

\markboth{}%
{Shell \MakeLowercase{\textit{et al.}}: Bare Demo of IEEEtran.cls for IEEE Journals}
%



\maketitle

\begin{abstract}
This paper studies the problem of recursively estimating the weighted adjacency matrix of a network out of a temporal sequence of binary-valued observations. 
The observation sequence is generated from nonlinear networked dynamics in which agents exchange and display binary outputs. 
Sufficient conditions are given to ensure stability of the observation sequence and identifiability of the system parameters. 
It is shown that stability and identifiability can be guaranteed under the assumption of independent standard Gaussian disturbances.
Via a maximum likelihood approach, the estimation problem is transformed into an optimization problem, and it is verified that its solution is the true parameter vector under the independent standard Gaussian assumption.
A recursive algorithm for the estimation problem is then proposed based on stochastic approximation techniques. 
Its strong consistency is established and convergence rate analyzed.
Finally, numerical simulations are conducted to illustrate the results and to show that the proposed algorithm is insensitive to small unmodeled factors.
\end{abstract}

{}

\IEEEpeerreviewmaketitle

\section{Introduction}\label{introduction}

In multiple scientific disciplines, network estimation, i.e., inferring underlying relationships between entities from static or dynamical data, is of great significance. 
For example, by estimating traffic volume between all pairs of nodes in a network from traffic flow, known as network tomography, traffic engineers can design new links to avoid network congestion \cite{vardi1996network}. 
Theoretical modeling and empirical verification of gene regulatory networks can enhance our understanding of diseases and development \cite{akutsu2018algorithms}. 
Lastly, inferring social structures such as friendship and influence can help to analyze and predict collective {behaviors} in complex social networks \cite{brugere2018network}.

An open problem in network estimation is to recursively estimate underlying networks based on quantized observations, despite extensive research efforts on how to reconstruct underlying networks from observing ordinary states or outputs, such as in graph signal processing \cite{ortega2018graph} and network inference for nonlinear systems \cite{timme2014revealing}.
Recursive algorithms \cite{ljung1983theory} are of great importance for identification of networked systems. 
They can be used for online tasks, such as adaptive control and decision-making, and thus have attracted much interest in the control community. 
More attention has been paid, however, on batch algorithms for network estimation, e.g., \cite{mateos2019connecting, ortega2018graph}. 
Quantized data are ubiquitous across domains, for example, active/inactive states of a gene \cite{akutsu2018algorithms}, ordinal rating of an individual \cite{durlauf2010social}, and failure conditions of an infrastructure \cite{baingana2014proximal}. 
Network estimation problems based on quantized time-series data need to be investigated in more depth with rigorous performance analysis \cite{mateos2019connecting, akutsu2018algorithms}.

The area of quantized identification, i.e., parameter estimation based on quantized data, has developed rapidly for the last decades \cite{wang2010system, marelli2013identification, bottegal2017new}. 
Many methods require either the design of input signals \cite{wang2010system} or of quantizers \cite{marelli2013identification,zhao2017recursive}. 
But these are different from our setting, because when estimating adjacency matrices from networked dynamics, there may be no possibility for imposing control inputs, and quantizers may be unknown components that cannot be designed. 

This paper studies a recursive network estimation problem based on binary-valued data, which is a special but crucial case of general quantized data. 
The binary data are generated from nonlinear networked dynamics, in which agents only exchange and display binary outputs. 
The nonlinear dynamics and limited observation information make the estimation problem hard. 
In order to solve the network estimation problem, we follow a maximum likelihood approach and propose a novel asymptotically consistent recursive algorithm.

\subsection{Motivating Examples}\label{motivation}

Dynamics with binary-valued observations can be encountered in a variety of domains. Here we present two {motivating} examples. 

\subsubsection*{Example 1. (Boolean Networks with Perturbation)}\label{motivatingExamplesA}
~\\
Boolean networks (BNs), first proposed by Kauffman \cite{kauffman1969metabolic}, have been extensively studied in many disciplines, including system biology \cite{akutsu2018algorithms}, physics \cite{aldana2003boolean}, and control theory \cite{cheng2010analysis}. 
BNs, where nodes have two states, representing active/inactive or ON/OFF, can be used to describe genetic regulatory networks, neural networks, disordered systems in statistical mechanics, and so on. 
To capture the intrinsic stochastic properties of these dynamics, researchers proposed various random versions of BNs \cite{akutsu2018algorithms}. 
Among these models, a particular one for network dynamics analyzed in \cite{aldana2003boolean, huepe2002dynamical}, can be mathematically described as follows. 

Consider a network consisting of $n$ nodes, $\mathcal{V} = \{1, \dots, n\}$, with an adjacency matrix $A=(a_{ij})$ capturing their relationships. Let $\bar{S}_k$ be the state vector at time $k \ge 0$. Node $i$ has state $\bar{S}_{k,i} \in \{1, -1\}$, $i \in \mathcal{V}$, and updates according to the averaged sum of its neighbors:
\begin{equation}\label{exampleAEquation}
\bar{S}_{k+1,i} = 
\begin{cases}
f\left(\sum_{j=1}^K a_{i,i_j} \bar{S}_{k, i_j}\right)  ~~\text{with probability } 1 - \eta,\\
-f\left(\sum_{j=1}^K a_{i,i_j} \bar{S}_{k, i_j}\right)  \text{with probability } \eta,
\end{cases}
\end{equation}
where for $x \in \mathbb{R}$, $f(x) = 1$ if $x \ge 0$ and $f(x) = -1$ otherwise, $i_j$ is a neighbor of $i$, $K$ is the total number of neighbors for every node, and $\eta \in (0, 1/2)$ is a constant. System \eqref{exampleAEquation} is an example of {a BN} with perturbation (BNp), where $\eta$ measures the intensity of the perturbation. 
Note that the function $f(\cdot)$ in \eqref{exampleAEquation} is a special case of Boolean threshold functions, which can be used to represent many Boolean functions \cite{akutsu2018algorithms}. 
Note also that \eqref{exampleAEquation} is related to another class of BNs called restricted BNs \cite{ouyang2014learning}. 

The identification of BNs is a significant issue, because underlying relations between nodes, either logical or parametric representations of the BNs. can be used for prediction and decision-making.
For \eqref{exampleAEquation}, the question is whether it is possible to estimate the adjacency matrix $A = (a_{ij})$ out of a temporal sequence of observation data $\{\bar{S}_k\}$.

\subsubsection*{Example 2. (Binary Choice Models of Social Interactions)}\label{motivatingExamplesB}
Social interactions shape behaviors of individuals. Numerous interactive decisions are binary, for example, voting and striking. 
As a result, lots of mathematical models have been proposed in order to analyze individual binary choices in social interactions \cite{durlauf2010social}. A binary-choice population process, whose update rule is related to a threshold function as in \eqref{exampleAEquation}, was studied in \cite{blume2003equilibrium} and described next. 

Each agent $i \in \{1, \dots, n\}$ has a binary state at time $k \ge 0$, $\bar{S}_{k,i}\in \{1, -1\}$, and updates to maximize a random utility function $V_{k,i}$, which depends on its neighbors' states:
\begin{align*}
V_{k, i}(s) 
&:= h_i s - \sum_{j=1}^n a_{ij} (s - \bar{S}_{k,j})^2  + \varepsilon_{k, i}(s)\\
&= h_i s + 2 s \sum_{j=1}^n a_{ij} \bar{S}_{k,j} + \varepsilon_{k, i}(s) - 2 \sum_{j=1}^n a_{ij},
\end{align*}
where $s$ takes value in $\{1, -1\}$, $h_i$ is a private preference, $a_{ij}$ is the conformity effect of $j$ on $i$, which can be {either positive or} negative, and $\{\varepsilon_{k, i}(1), k \ge 0, 1 \le i \le n\}$, $\{\varepsilon_{k, i}(-1), k \ge 0, 1 \le i \le n\}$ are mutually independent random sequences, both independent and identically distributed (i.i.d.). Hence according to this utility function, the probability that agent $i$ takes choice $1$ at time $k+1$ is
\begin{align}\nonumber
&P\{\bar{S}_{k+1,i} = 1\}\\\nonumber 
&= P\{V_{k, i}(1) - V_{k, i}(-1) \ge 0\}\\\nonumber
&= P\{\varepsilon_{k, i}(1) - \varepsilon_{k, i}(-1) \ge -2h_i - 4\sum_{j=1}^n a_{ij} \bar{S}_{k,j}\}\\\label{exampleBEquation}
&= 1 - F(-2h_i - 4\sum_{j=1}^n a_{ij} \bar{S}_{k,j}),
\end{align}
where $F(\cdot)$ is the cumulative distribution function of $\varepsilon_{k, i}(1) - \varepsilon_{k, i}(-1)$. It is of interest whether we can recover conformity relationships between agents based on observed binary actions.

\subsection{Related Work}

For the identification of BNs, most work has been on estimating logical interrelations or Boolean functions of deterministic BNs \cite{cheng2011identification, akutsu2018algorithms}.
The problem considered in this paper, however, is on estimating the adjacency matrix and determining the Boolean threshold functions in a BNp. 
In \cite{marshall2007inference}, an estimation procedure for BNps from temporal data sequence was proposed based on a transition counting matrix and the optimal selection of input nodes. 
The authors studied the estimation problem of restricted BNs in \cite{ouyang2014learning, higa2011constraint}, but did not present a rigorous performance analysis. 
Additionally, \cite{melkman2017identifying, akutsu2018identification} investigated the inferring of Boolean threshold functions for probabilistic BNs, but data samples were assumed to be independent instead of taken from a time series.

Results on identification for binary choice models in the field of econometrics can be found in \cite{blume2011identification}, which established sufficient conditions for identifiability.
Many estimation methods and their asymptotic properties have, however, been considered under static games and by letting the network size tend to infinity \cite{menzel2015inference,yang2018tobit}. 

A related class of binary state models is cascading dynamics, where node states are interpreted as functioning or failing, and the failure condition is assumed to be absorbing. Research on estimating networks from cascading dynamics can be found in \cite{baingana2014proximal}. Besides, the authors of \cite{wu2018estimating} studied the network estimation problem for a discuss-then-vote model, in which individuals display a discrete voting choice at the end of each discussion, but they still exchange continuous states during the discussion. 

In the literature of quantized identification, there are multiple methods not relying on the design of inputs or quantizers. 
For example, the maximum likelihood method was used in \cite{godoy2011identification, aguero2017based, risuleo2019identification, marelli2013identification}. 
An online algorithm based on the expectation-maximization (EM) algorithm and quasi-Newton method for autoregressive moving average (ARMA) models with quantized outputs was studied in \cite{marelli2013identification}{.} 
To achieve the best performance, quantizers need to be known and adaptive. 
In \cite{godoy2011identification, aguero2017based}, the EM algorithm was used to optimize the likelihood function, while in \cite{risuleo2019identification} a variational approximation approach was utilized. 
Additionally, Bayesian frameworks were applied in, e.g., \cite{bottegal2017new}. 
The authors of \cite{wigren1998adaptive} proposed an algorithm based on the recursive prediction error method to estimate the linear part of Wiener systems, which can be used to deal with quantized output models, {but both quantizers and the range of parameters were assumed to be known}. 
A least-squares algorithm was developed in \cite{jafari2012convergence} to recursively estimate finite impulse response systems.
For the theoretical results the authors assumed that the inputs have a positive-measure support and that the threshold is known.

\subsection{Contributions}

This paper studies a recursive network estimation problem based on binary data.
More specifically, we recursively estimate the weighted adjacency matrix of a network out of a temporal sequence of binary observations.
The observation sequence is generated from nonlinear networked dynamics in which agents exchange and display binary outputs.

Our contributions are summed up as follows.

1. We tackle the recursive network estimation problem for a nonlinear dynamic network based on binary data, by proposing a strongly consistent estimation algorithm.
Different from existing batch algorithms, the recursive algorithm can be applied to online tasks.

2. We show stability of the observation sequence and investigate identifiability under different disturbance assumptions. 
Stability and identifiability can be guaranteed under the assumption of independent standard Gaussian disturbances. In addition, it is shown that the Gaussian assumption can be relaxed if more conditions are imposed on the adjacency matrix, and that identifiability may not hold if the disturbances are discrete random variables.

3. We propose an optimization function based on the maximum likelihood estimators. It is verified that under the assumption of independent standard Gaussian disturbances that function is strictly concave and has the true parameter vector as its unique maximum.
Our recursive algorithm is shown to seek this maximum, by using stochastic approximation techniques. 
The algorithm is verified to be strongly consistent, and its convergence rate is estimated. 

The differences of this paper from the conference version \cite{xing2019network} are that we present motivating examples, give rigorous proofs of the theorems, analyze the convergence rate of the algorithm, and show numerical simulations to demonstrate properties of the algorithm.

\subsection{Outline}
The remainder of this paper is organized as follows.
In Section \ref{problem formulation}, the network estimation problem is formulated.
Stability of the observation sequence and identifiability of the system parameters are studied in Section \ref{model analysis}. 
In Section \ref{network estimation algorithm}, we propose the network estimation algorithm, and then analyze its strong consistency and convergence rate. 
Section \ref{numerical simulations} presents numerical simulations showing that the proposed algorithm is robust to small unmodeled dynamics, and Section \ref{conclusion} concludes the paper. To keep the paper fluent, some proofs are postponed to appendices.

By boldfaced lower-case or Greek letters we denote column vectors, and by upper-case letters we denote matrices and random vectors. We use $\mathbb{R}$, $\mathbb{R}^n$, $\mathbb{R}^{n\times m}$, and $\|\cdot\|$ to represent the set of real numbers, the $n$-dimensional Euclidean space, the set of $n\times m$ real matrices, and the Euclidean norm of a vector, respectively. Let $\bm{0}_n$, $\bm{1}_n$, and $\bm{e}_i$ be the $n$-dimensional all-zero vector, the $n$-dimensional all-one vector, and the unit vector with $i$-th entry being one. 

By $a_i$ and $\bm{a}_{i:j}$ we denote the $i$-th entry of vector $\bm{a}$ and its sub-vector $(a_i, a_{i+1}, \dots, a_j)^T$. For a matrix $A \in \mathbb{R}^{n\times m}$, $a_{ij}$, $A_i$, and $A^T$ are used to represent its entry $(i,j)$, $i$-th row, and transpose. Define $\text{vec}(A) := (a_{11}~a_{12}~\cdots~a_{1n}~a_{21}~\cdots~a_{2m}~\cdots~a_{nm})^T$. Denote the absolute value of $x \in \mathbb{R}$ by $|x|$, $|\bm{a}| := (|a_1|, \dots, |a_n|)^T$, and $|A| := (|a_{ij}|)$. A matrix $A\in\mathbb{R}^{n\times n}$ is called stochastic if $A \bm{1}_n = \bm{1}_n$, and called absolutely stochastic if $|A| \bm{1}_n = \bm{1}_n$.

Let $E\{X_k\}$, $X_{k, i}$, and $X_{k, i:j}$ be the expectation, the $i$-th entry, and the sub-vector $(X_{k, i}, X_{k, i+1}, \dots, X_{k, j})^T$ of {{a} random vector $X_k$, $k \ge 0$. 
For $a, b \in \mathbb{R}$, denote $a \vee b := \max\{a, b\}$ and $a \wedge b := \min \{a, b\}$. Define $\mathcal{S}^n$ as the Descartes product $\times_{i = 1}^n \mathcal{U}_i$, where $\mathcal{U}_i = \{1, 0\}$, $1 \le i \le n$. $\mathbb{I}_{[\text{inequality}]}$ is the indicator function equal to $1$, if the inequality holds, and equal to $0$ otherwise. 
The gradient and the Hessian of {{$f(\bm{x})$} with respect to $\bm{x}$ are denoted by $\nabla_{\bm{x}}f(\bm{x})$ and $\nabla^2_{\bm{x}}f(\bm{x})$, respectively. 
{For two sequences $\{a_k\}$ and $\{b_k\}$ with $b_k \not= 0$, $k \ge 1$, $a_k = O(b_k)$ means that $\lim_{k \to \infty} |a_k/b_k| \le C$ for some positive number $C$, and $a_k = o(b_k)$ means that $\lim_{k \to \infty} |a_k/b_k| = 0$.}

For a homogeneous and finite-state Markov chain $\{X_k\}$ in a state space $\Omega$, the transition probability from $x$ to $y$ is $P(x, y) := P\{X_1 = y|X_0 = x\}$, and the $k$-step transition probability from $x$ to $y$ is $P^k(x, y) := P\{X_k = y|X_0 = x\}$, $\forall x, y \in \Omega$. 
We say that $y$ is reachable from $x$, if there exists $k \ge 1$ such that $P^k(x, y) > 0$. 
The Markov chain is said to be irreducible, if $y$ is reachable from $x$ for all $x, y \in \Omega$. 
The greatest common divisor of set $\{k \ge 1: P^k(x, x) > 0\}$ is called the period of $x$, denoted by $d(x)$. 
The Markov chain is aperiodic if $d(x) = 1$ for all $x \in \Omega$. 
{{A probability distribution $\pi$ on $\Omega$ is referred to} as a stationary distribution of $\{X_k\}$, if $\forall y \in \Omega$, $\pi(y) = \sum_{x \in \Omega} \pi(x) P(x, y)$.

\section{Problem Formulation}\label{problem formulation}

\subsection{Problem}

In the sequel, suppose that the network size $n \ge 2$. 
The considered dynamics with binary observations in this paper is as follows:
\begin{equation}\label{binary}
\begin{aligned}
Y_{k+1}&= A S_k + D_k,\\
S_k &= \mathcal{Q}(Y_k, \bm{c}),
\end{aligned}
\end{equation}
where $k \ge 0$, $Y_k = (Y_{k, 1}, \dots, Y_{k, n})^T$, $D_k =(D_{k, 1}, \dots,$ $D_{k,n})^T$, $S_k = (S_{k, 1}, \dots, S_{k, n})^T$ are the inner state, the disturbance, and the observation vector at time $k$, respectively. 
$A \in \mathbb{R}^{n\times n}$ is the weighted adjacency matrix, and $\bm{c} = (c_{1}, \dots, c_{n})^T \in \mathbb{R}^n$ is the unknown quantization threshold vector. 
$\mathcal{Q}(Y_k, \bm{c}) := (\mathbb{I}_{[Y_{k, 1} \ge c_1]}, \dots, \mathbb{I}_{[Y_{k, n} \ge c_n]})^T$ is the quantizer. See Fig. \ref{struc} for an illustration of this system.

For the weighted adjacency matrix $A$, we do not assume that it is strongly connected or that its row sums are equal to one. 
Negative weights, representing antagonistic relationships, are also permitted. 
A detailed discussion on assumptions for $A$ is in Section \ref{sec:identifiability}.

\begin{figure}
\centering
\includegraphics[scale=0.65]{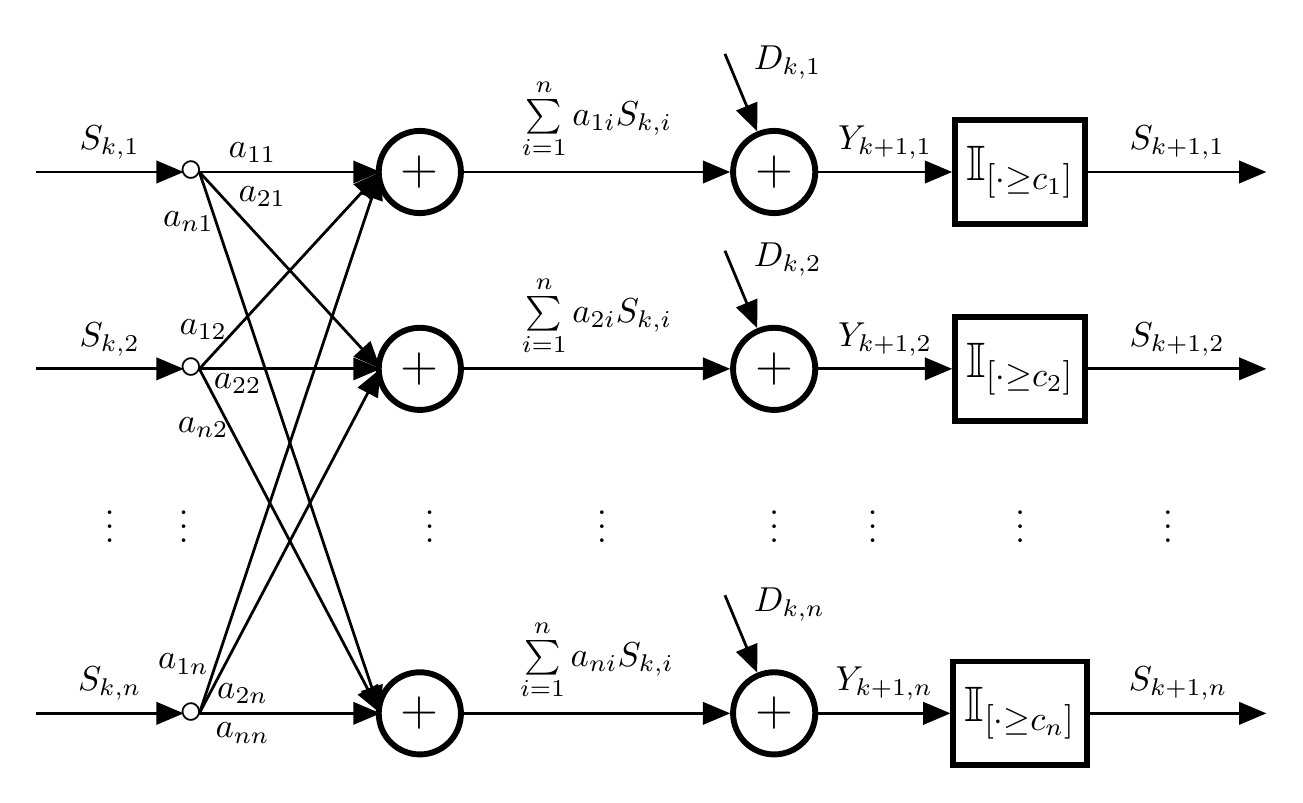}
\caption{\label{struc}The problem considered in this paper is how to recursively estimate the weighted adjacency matrix $A = (a_{ij})$ and the quantization threshold $\bm{c}$ out of observations $\{S_k\}$.}
\end{figure}

The problem considered in this paper is to recursively estimate the weighted adjacency matrix $A$ and the quantization threshold vector $\bm{c}$ out of the observation sequence $\{S_k\}$. This problem actually consists of two key questions: is it possible to estimate the parameters only from binary-valued observations? If so, how to recursively estimate the parameters? We investigate these two questions in Section \ref{model analysis} and \ref{network estimation algorithm}, respectively.

\subsection{Motivating Examples Revisited}\label{quickRevisit}

We briefly revisit the motivating examples to show that they fit into the model \eqref{binary}.

For Example $1$, note that although we follow the convention of Boolean networks in \eqref{binary} and define the states as $\{1,0\}$, it can be transformed to a system with observation states $\{1,-1\}$: $\bar{S}_k = 2S_k - \bm{1}_n$, and
\begin{equation}\label{binaryBar}
\begin{aligned}
\bar{Y}_{k+1}&= A \bar{S}_k + \bar{D}_k,\\
\bar{S}_k &= \bar{\mathcal{Q}}(\bar{Y}_k, \bar{\bm{c}}),
\end{aligned}
\end{equation}
where $\bar{Y}_k = 2Y_k - A\bm{1}_n$, $\bar{D}_k = 2D_k$, $\bar{\bm{c}} = 2 \bm{c} - A \bm{1}_n$, and $\bar{\mathcal{Q}}(\bar{Y}_k, \bar{\bm{c}}) = 2\mathcal{Q}(\bar{Y}_k, \bar{\bm{c}}) - \bm{1}_n = 2\mathcal{Q}(Y_k, \bm{c}) - \bm{1}_n$. 
In \eqref{binaryBar}, $Y_k$, $D_k$, and $\bm{c}$ have been changed accordingly to the observation transformation. 
Hence, \eqref{exampleAEquation} is equivalent to \eqref{binaryBar} with disturbance $\bar{D}_{k,i}$ and threshold $\bar{c}_i$ such that $P\{\bar{D}_{k,i} \ge \bar{c}_i - A_i\bar{\bm{s}}\} = (1 - \eta) \mathbb{I}_{[A_i\bar{\bm{s}} \ge 0]} + \eta\mathbb{I}_{[A_i\bar{\bm{s}} < 0]}$ for $\bar{\bm{s}} \in \{1,-1\}^n$, $1 \le i \le n$, and $k \ge 0$. 
As a matter of fact, as discussed in Section \ref{model analysis} (Assumption \ref{noiseAsmpPPP}), $\{\bar{D}_{k,i}\}$ can be a sequence of i.i.d. discrete random variables satisfying 
\begin{equation*}
\bar{D}_{k,i} =
\begin{cases}
\bar{d}_{i1} & \text{with probability } \eta,\\
\bar{d}_{i2} & \text{with probability } 1 - 2\eta,\\
\bar{d}_{i3} & \text{with probability } \eta,
\end{cases}
\end{equation*}
with $\eta \in (0, 1/2)$, $\bar{d}_{i1} < \bar{c}_i - A_i\bar{\bm{s}}$, $\bar{d}_{i3} \ge \bar{c}_i - A_i\bar{\bm{s}}$, for all $\bar{\bm{s}} \in \{1,-1\}^n$, and $\bar{d}_{i2} = \bar{c}_i$. 
In this way, for $\bar{\bm{s}}$ such that $A_i\bar{\bm{s}} \ge 0$, {it holds that} $\bar{c}_i - A_i\bar{\bm{s}} \le \bar{c}_i = \bar{d}_{i2} \le \bar{d}_{i3}$, so $P\{\bar{D}_{k,i} \ge \bar{c}_i - A_i\bar{\bm{s}}\} = 1 - \eta$. On the other hand, $P\{\bar{D}_{k,i} \ge \bar{c}_i - A_i\bar{\bm{s}}\} = \eta$ for $\bar{\bm{s}}$ such that $A_i\bar{\bm{s}} < 0$. It can also be observed that if $\eta=0$, then \eqref{binaryBar} is deterministic. 

We have the same state-space transformation for Example $2$ as above, so \eqref{exampleBEquation} is equivalent to \eqref{binaryBar} with $\bar{c}_i = -\frac12 h_i$ and $\bar{D}_{k,i} = \frac14 (\varepsilon_{k,i}(1) - \varepsilon_{k,i}(-1))$, for $1 \le i \le n$, $k \ge 0$.

\section{Model Analysis}\label{model analysis}
In this section, we study stability of the observation sequence and identifiability of the system parameters, and provide sufficient conditions such that the network estimation problem is well-posed.

\subsection{Stability of Observation Sequences}
As in \eqref{binary}, the observation sequence $\{S_k, k \ge 0\}$ is a Markov chain with finite states. 
The existence of stationary distributions is a significant aspect of stochastic stability of Markov chains \cite{meyn2012markov}, and we have a straightforward result under the following assumption.

\begin{asmp}\label{noiseAsmp}(Disturbance)
The disturbances of \eqref{binary} satisfy that \\
i) $\{D_{k, i}, k \ge 0\}$ are sequences of i.i.d. random variables, mutually independent, and independent of $S_0$, $1 \le i \le n$;\\
ii) both $P\{D_{k, i} \ge c_i + |A_i|\bm{1}_n\} > 0$ and $P\{D_{k, i} < c_i - |A_i| \bm{1}_n\} > 0$ hold for $1 \le i \le n$, $k \ge 0$.
\end{asmp}

\begin{thm}\label{MC}(Stability)\\
Suppose that Assumption \ref{noiseAsmp} holds, then Markov chain $\{S_k\}$ is irreducible and aperiodic. Moreover, $P(\bm{u}, \bm{s}) > 0$ holds for any $\bm{u}, \bm{s} \in \mathcal{S}^{n}$. Hence, $\{S_k\}$ converges in distribution, from any initial condition, to a unique stationary distribution $\pi$ on $\mathcal{S}^n$ with $\pi(\bm{s}) > 0$, $\forall \bm{s} \in \mathcal{S}^n$.
\end{thm}

\begin{prof}\textup{
The conclusion follows from directly computing the transition probabilities of $\{S_k\}$, which is similar to the proof of Theorem $1$ {in the conference version of this paper} \cite{xing2019network}.}
\hfill$\Box$
\end{prof}

\begin{rmk}
Theorem \ref{MC} provides a sufficient condition for the irreducible and aperiodic properties of $\{S_k\}$, and Assumption \ref{noiseAsmp} is strong enough so that we do not need extra assumptions for the weighted adjacency matrix $A$. In fact, the {behaviors} of System \eqref{binary} and related models {have} been extensively studied in different disciplines, including stability, attractor analysis, and so on (see e.g. \cite{aldana2003boolean, huepe2002dynamical, higa2011constraint, blume2003equilibrium}). 
Nevertheless, we present this theorem to show that the observation sequence can exhibit sufficient diversity, as long as the disturbance can surpass the influence of others {on} an agent, making this agent display a different action from its previous one. The diversity is necessary for a successful estimation of the weighted adjacency matrix, playing a crucial role as persistent excitement does \cite{ljung1987system}.
\end{rmk}

Define $\tilde{S}_k := (S_k^T~ S_{k - 1}^T)^T$, $k \ge 1$. This auxiliary chain is critical for our estimation. Note that $\{\tilde{S}_k\}$ taking values in $\mathcal{S}^{2n}$ is also a Markov chain. For $k \ge 1$ and $\bm{s}_{k - 1}, \bm{s}_k$, $\bm{s}_{k + 1} \in \mathcal{S}^n$, {it} holds that
\begin{equation}\label{tildeS}
\begin{aligned}
&P\left\{\tilde{S}_{k + 1} = \begin{pmatrix} \bm{s}_{k + 1} \\ \bm{s}_k \end{pmatrix} \Big| \tilde{S}_{k} = \begin{pmatrix} \bm{s}_{k} \\ \bm{s}_{k - 1} \end{pmatrix} \right\} \\
&= P\{S_{k + 1} = \bm{s}_{k + 1}|S_k = \bm{s}_k\}.
\end{aligned}
\end{equation}
So $\{\tilde{S}_k\}$ is aperiodic. For states $(\bm{s}^T~\bm{u}^T)^T, (\bm{x}^T~\bm{y}^T)^T \in \mathcal{S}^{2n}$, since $\{S_k\}$ is irreducible, there exists $k \ge 1$ such that $P^k(\bm{x}, \bm{u}) > 0$. 
Moreover, from {Theorem} \ref{noiseAsmp}, $P(\bm{u}, \bm{s}) > 0$ holds. Hence it follows from \eqref{tildeS} that 
\[
P\left\{\tilde{S}_{k + 1} = \begin{pmatrix} \bm{s} \\ \bm{u} \end{pmatrix} \Big| \tilde{S}_{0} = \begin{pmatrix} \bm{x} \\ \bm{y} \end{pmatrix} \right\} > 0,
\]
which implies that $\{\tilde{S}_k\}$ is also irreducible, and further we have the following result:

\begin{thm}\label{MC2}(Stability of the auxiliary chain)\\
Suppose that Assumption \ref{noiseAsmp} holds, then Markov chain $\{\tilde{S}_k\}$ is irreducible and aperiodic. Hence, it converges in distribution, from any initial condition, to a unique stationary distribution $\tilde{\pi}$ on $\mathcal{S}^{2n}$ with $\tilde{\pi}(\tilde{\bm{s}}) > 0$, $\forall \tilde{\bm{s}} \in \mathcal{S}^{2n}$.
\end{thm}

The next lemma illustrates the relation between $\{S_k\}$ and the stationary distribution of $\{\tilde{S}_k\}$, which is crucial for our theoretical results but also has its own intuitive meaning.

\begin{lem}\label{lem1}
Suppose that Assumption \ref{noiseAsmp} holds, and $\tilde{S}$ is subject to the stationary distribution of $\{\tilde{S}_k\}$. Then 
\begin{equation*}
P\{\tilde{S}_{1:n} = \tilde{\bm{s}}_{1:n} | \tilde{S}_{n+1:2n} = \tilde{\bm{s}}_{n+1:2n}\} = P(\tilde{\bm{s}}_{n+1:2n}, \tilde{\bm{s}}_{1:n}),
\end{equation*}
$\forall \tilde{\bm{s}} \in \mathcal{S}^{2n}$, where $P(\cdot, \cdot)$ is the probability transition matrix of $\{S_k\}$.
\end{lem}

\begin{proof}
See Appendix \ref{AppendixA}.
\end{proof}

\begin{rmk}
This lemma indicates that the conditional probability of the event $\{\tilde{S}_{1:n} = \tilde{\bm{s}}_{1:n}\}$ given $\{\tilde{S}_{n+1:2n} = \tilde{\bm{s}}_{n+1:2n}\}$ is the same as the transition probability of $\{S_k\}$ from $\tilde{\bm{s}}_{n+1:2n}$ to $\tilde{\bm{s}}_{1:n}$, $\forall \tilde{\bm{s}} \in \mathcal{S}^{2n}$. This accords with the definition of $\{\tilde{S}_k\}$.
\end{rmk}

In most parts of this paper, we adopt the following standard Gaussian assumption for disturbances, since a maximum likelihood approach is utilized. The reason for fixing the variance to be {one} is discussed in the next section.

\begingroup
\def\theasmp{\ref{noiseAsmp}$^\prime$}
\addtocounter{asmp}{-1}
\begin{asmp}\label{noiseAsmpP}
The disturbances of \eqref{binary} satisfy that \\
i) $\{D_{k, i}, k \ge 0\}$ are sequences of i.i.d. random variables, mutually independent, and independent of $S_0$, $1 \le i \le n$;\\
ii) $D_{k,i} \sim \mathcal{N}(0, 1)$, for $1 \le i \le n$, $k \ge 0$, where $\mathcal{N}(0, 1)$ represents the standard Gaussian distribution.
\end{asmp}
\endgroup

\begin{rmk}
Since the probability density function of the standard Gaussian {distribution} is positive on $\mathbb{R}$, ii) of Assumption \ref{noiseAsmpP} implies {ii)} of Assumption \ref{noiseAsmp}. 
Thus Theorems \ref{MC}, \ref{MC2}, and Lemma \ref{lem1} still hold under Assumption \ref{noiseAsmpP}. 
But to avoid repetition, we summarize them as Theorems \ref{MCprime}, \ref{MC2prime}, and Lemma \ref{lem1prime} in Appendix \ref{AppendixA}.
\end{rmk}

\subsection{Identifiability}\label{sec:identifiability}

We have shown in the preceding subsection that certain conditions ensure diverse information for estimation, but before proposing the estimation algorithm, we need to study identifiability of the parameters of \eqref{binary}. 
This is because consistent estimators cannot exist if the parameters are not identifiable.

For \eqref{binary}, since the only available observation is $\{S_k\}$ with Markov property, we define the identifiability from a statistical perspective. Because a finite-state Markov chain is uniquely defined by its transition probability matrix, the following definition is introduced, where we denote the parameter vector by $\theta := \text{vec}\big\{(A~\bm{c})\big\}$ with $(A~\bm{c}) \in \mathbb{R}^{n\times (n+1)}$ {and the transition probability matrix of $\{S_k\}$ by $P_{\theta}$, emphasizing the dependence of $\{S_k\}$ on $\theta$.}

\begin{defn}
The parameters of \eqref{binary} are identifiable, if $P_{\theta} = P_{\hat{\theta}}$ implies $\theta = \hat{\theta}$, $\forall \theta, \hat{\theta} \in \mathbb{R}^{{n\cdot(n+1)}}$.
\end{defn}

\begin{rmk}
The identifiability is defined differently in various disciplines, for example, in the sense of input-output sequences in the control community \cite{ljung1987system}, in a combinatorial way in system biology \cite{akutsu2018algorithms}, and in terms of distributions in statistics \cite{van2000asymptotic}. 
Our definition follows the last one because time-series data with Markov property is of consideration and there is no explicit input-output relations. 
These definitions, however, all demonstrate the same idea that if a system is uniquely defined by parameters, then it will generate distinctive observations. 
\end{rmk}

Under Assumption \ref{noiseAsmpP} we have the following identifiability result for \eqref{binary}.

\begin{thm}(Identifiability)\label{ident}
Suppose that Assumption \ref{noiseAsmpP} holds, then {parameters of} \eqref{binary} are identifiable.
\end{thm}

\begin{proof}
See the proof of Theorem $3$ in the conference version of this paper \cite{xing2019network}.
\end{proof}

\begin{rmk}
This theorem shows that under Assumption \ref{noiseAsmpP} the parameter vector $\theta$ can be uniquely determined if the probability transition matrix of $\{S_k\}$ is known. On the other hand, in order to tackle the estimation problem, sufficiently abundant observations are guaranteed by Theorem \ref{MC}.
\end{rmk}

This conclusion is similar to well-known ones in system identification with the presence of binary sensors (e.g. \cite{godoy2011identification}). 
In fact, it is easy to observe that when the variances of the Gaussian disturbances are not fixed, there exist a set of systems that define an identical observation process:

\begin{prop}\label{identP}
{Suppose that Assumption \ref{noiseAsmpP} holds}, then the following system defines the same transition probability matrix of the observation process as that of $\{S_k\}$ in \eqref{binary} ,
\begin{equation}\label{equivalentForm}
\begin{aligned}
\tilde{Y}_{k+1} &= \tilde{A} S_k + \tilde{D}_k,\\
S_{k} &= \mathcal{Q}(\tilde{Y}_k, \tilde{\bm{c}}),
\end{aligned}
\end{equation}
where {for} $B = \textup{diag}(b_1, \dots, b_n)$, a diagonal matrix with non-zero diagonal entries, $\tilde{A} = B^{-1} A$, $\tilde{D}_k = B^{-1} D_k$, $\mathcal{Q}(\tilde{Y}_k, \tilde{\bm{c}}) = (\mathbb{I}_{[\tilde{Y}_{k, 1} > \tilde{c}_1]}, \dots, \mathbb{I}_{[\tilde{Y}_{k, n} > \tilde{c}_n]})^T$, and $\tilde{c}_i = b_i^{-1} c_i$, for $k \ge 0$, $1 \le i \le n$. 
\end{prop}

\begin{rmk}\label{IDrmk}
{If $\forall 1 \le i \le n, \exists j \not= i$, $a_{ij} \not =0$, then let $b_i = |A_i| \bm{1}_n \not = 0$, $1 \le i \le n$.}
Hence $\tilde{A}$ is absolutely row stochastic, and $\tilde{D}_{k, i}$, $1 \le i \le n$, become Gaussian random variables with different variances. 
That is to say, if $A$ is assumed to be absolutely stochastic, then the variances of $\tilde{D}_{k, i}$ are unnecessarily assumed to be one and known.
\end{rmk}

Another interesting question is whether parameters of \eqref{binary} are identifiable under discrete disturbances as in Example $1$. 

\begingroup
\def\theasmp{\ref{noiseAsmp}$^{\prime\prime}$}
\addtocounter{asmp}{-1}
\begin{asmp}\label{noiseAsmpPPP}
The disturbances of \eqref{binary} satisfy that \\
i) $\{D_{k, i}, k \ge 0\}$ are sequences of i.i.d. random variables, mutually independent, and independent of $S_0$, $1 \le i \le n$; \\
ii) for $1 \le i \le n$ and $k \ge 0$, 
\begin{equation*}
D_{k,i} =
\begin{cases}
d_{i1} & \text{with probability } \eta,\\
d_{i2} & \text{with probability } 1 - 2\eta,\\
d_{i3} & \text{with probability } \eta,
\end{cases}
\end{equation*}
where $\eta \in (0,1/2)$, and $d_{i1}$, $d_{i2}$, $d_{i3} \in \mathbb{R}$ are such that $d_{i1} < c_i - A_i \bm{s}$ and $d_{i3} \ge c_i -A_i \bm{s}$, $\forall \bm{s} \in \mathcal{S}^n$, and $d_{i2} = c_i - \frac12 A_i \bm{1}_n$.
\end{asmp}
\endgroup

\begin{rmk}
Note that ii) of Assumption \ref{noiseAsmpPPP} in this case is a weaker condition than that in Assumption \ref{noiseAsmp}. Also, ii) of Assumption \ref{noiseAsmpPPP} guarantees the condition of disturbances for Example $1$ by noticing the transformation used to obtain \eqref{binaryBar}.
\end{rmk}

Under Assumption \ref{noiseAsmpPPP}, unfortunately, {parameters of} \eqref{binary} {are} not identifiable, if {they} take real numbers. 
This can be seen from an intuitive observation that discrete random variables are not that ``sensitive'', so slightly modifying $A_i$ does not change the transition probability of \eqref{binary}.

\begingroup
\def\thethm{\ref{ident}$^{\prime}$}
\addtocounter{thm}{-1}
\begin{thm}\label{unident}
{Suppose that Assumption \ref{noiseAsmpPPP} holds. If $\forall 1 \le i \le n, \exists j \not= i$, $a_{ij} \not =0$,} then parameters of \eqref{binary} are not identifiable.
\end{thm}
\endgroup

\begin{proof}
See Appendix \ref{AppendixB}.
\end{proof}

\begin{rmk}
For the weighted adjacency matrix $A$, entry $(i,j)$ represents the influence of $j$ on $i$, so the assumption for $A$ in this theorem means that every agent has certain connections with others, but the graph may not necessarily be strongly connected.
It can be seen that this result still holds for parameters taking rational numbers, but it does not contradict with some existing identifiability results, since they assume that the adjacency matrix is integer-valued \cite{akutsu2018algorithms}. In fact, it again indicates that restricting assumptions are needed to establish well-posed identifiability, because of the limited binary-valued information.
\end{rmk}

\section{Network Estimation}\label{network estimation algorithm}

In this section, in order to tackle the recursive network estimation problem, first we prove that it is equivalent to seeking a unique maximum of an objective function, which is related to the stationary distribution of the observation sequence, under the assumption of independent standard Gaussian disturbances. 
The objective function, however, cannot be obtained directly, so an online algorithm based on stochastic approximation techniques is developed to solve the optimization task. 
Finally, asymptotic properties of the proposed algorithm {are} studied, including strong consistency and convergence rate.

\subsection{An Objective Function and Its Concavity}

Recall that $\theta = \text{vec}\big\{(A~\bm{c})\big\}$ is the parameter vector to be estimated, and further denote $\theta^{(i)} = (A_i~c_i)^T$. To avoid ambiguity, $\theta^* := \text{vec}\big\{(A^*~\bm{c}^*)\big\} = (((\theta^*)^{(1)})^T, \dots, (\theta^*)^{(n)})^T)^T$ is used to represent the true parameters. Given observation data $\{\bm{s}^k, 0 \le k \le T\}$, the log likelihood function is
\begin{align}\nonumber
&l(T ; \theta) \\\nonumber
&= \log P\{S_k = \bm{s}^k, 0 \le k \le T\} \\\nonumber
&= \log P\{S_0 = \bm{s}^0\} + \sum\limits_{1 \le k \le T} \log P\{S_k = \bm{s}^k | S_{k - 1} = \bm{s}^{k - 1}\}\\\label{gii}
&= \log P\{S_0 = \bm{s}^0\} + \sum\limits_{1 \le k \le T} \sum\limits_{1 \le i \le n} \log g_i(\tilde{\bm{s}}^k|\theta^{(i)}),
\end{align}
where $(\tilde{\bm{s}}^k)^T := ((\bm{s}^k)^T(\bm{s}^{k - 1})^T)$, and  
\begin{equation}\label{g_i}
g_i(\tilde{\bm{s}}|\theta^{(i)}) := (1 - \Phi(c_i - A_{i} \tilde{\bm{s}}_{n+1:2n}))^{\tilde{s}_i} \Phi(c_i - A_{i}\tilde{\bm{s}}_{n+1:2n})^{1 - \tilde{s}_i},
\end{equation}
{ for $\tilde{\bm{s}} \in \mathcal{S}^{2n}$ and $1 \le i \le n$.}
{Here $\Phi(x)$ represents the cumulative density function of the standard Gaussian distribution.}

For fixed $\theta$, $g_i(\tilde{\bm{s}}|\theta^{(i)})$ and $\nabla_{\theta^{(i)}} g_i(\tilde{\bm{s}}|\theta^{(i)})$ are bounded since $\tilde{\bm{s}}$ takes values in $\mathcal{S}^{2n}$. Thus, by ergodic properties of Markov chains (Theorem 17.1.7 in \cite{meyn2012markov}), the following equations hold for the chain $\{\tilde{S}_k\}$ and fixed $\theta$ a.s.:
\begin{align*}
& \lim_{T \to \infty} \frac1T \sum\nolimits_{1 \le k \le T} \sum\nolimits_{1 \le i \le n} \log g_i(\tilde{S}_k|\theta^{(i)}) \\
&= E \bigg\{\sum\nolimits_{1 \le i \le n} \log g_i(\tilde{S}|\theta^{(i)}) \bigg\},\\
&\lim_{T \to \infty} \frac1T \sum\nolimits_{1 \le k \le T} \sum\nolimits_{1 \le i \le n} \nabla_{\theta^{(i)}} \log g_i(\tilde{S}_k|\theta^{(i)}) \\
&= E \bigg\{\sum\nolimits_{1 \le i \le n} \nabla_{\theta^{(i)}} \log g_i(\tilde{S}|\theta^{(i)})\bigg\}, 
\end{align*}
where $\tilde{S}$ is subject to the stationary distribution of $\{\tilde{S}_k\}$.

Therefore, the function of $\theta$ 
\begin{equation} \label{objectiveFunction}
E \bigg\{\sum\nolimits_{1 \le i \le n} \log g_i(\tilde{S}|\theta^{(i)})\bigg\} 
\end{equation} 
will be used as an objective function to fulfill the estimation of $\theta^*$. It has a good property: 

\begin{thm}(Strict concavity of \eqref{objectiveFunction})\label{mainthm1}\\
Under Assumption \ref{noiseAsmpP}, the function \eqref{objectiveFunction} of $\theta$ is strictly concave over $\mathbb{R}^{{n\cdot(n+1)}}$, and the true parameter vector $\theta^*$ is its unique maximum point.
\end{thm}

\begin{proof}
See Appendix \ref{AppendixC}
\end{proof}

\begin{rmk}
This theorem is the key to establish the consistent estimation of the weighted adjacency matrix $A^*$, since it shows that $\theta^* = \textup{vec}\big\{(A^*~\bm{c}^*)\big\}$ can be obtained by optimizing \eqref{objectiveFunction}. Because of its significance, one of the future works of this paper is to generalize the disturbance assumption.
\end{rmk}

Therefore, our estimation task turns to seeking the unique maximum point of this function. However, $\tilde{S}$ cannot be directly obtained, so the observations $\{\tilde{S}_k\}$ are used to replace it. A stochastic approximation (SA) algorithm is introduced in next {subsection}, and it is verified that the true network can indeed be estimated by using the observation sequence.

\subsection{Network Estimation Algorithm}

We use the SA algorithm to deal with the estimation problem. For $1 \le i \le n$ and $k \ge 1$, denote
\begin{equation}\label{Ki}
K_{i}(\theta^{(i)}, \tilde{S}_{k}) := \nabla_{\theta^{(i)}} \log g_i(\tilde{S}_{k} | \theta^{(i)}), 
\end{equation}
\begin{equation}\label{K}
K(\theta, \tilde{S}_{k}) := (K_1(\theta^{(1)}, \tilde{S}_{k}), \dots, K_n(\theta^{(n)}, \tilde{S}_{k}))^T,
\end{equation}
where $\theta^T = ((\theta^{(1)})^T, \dots, (\theta^{(n)})^T)$, and $g_i$ is defined in \eqref{g_i}.

The estimation algorithm is as follows:
\begin{equation}\label{IdenAlg}
\theta_{k + 1} = \theta_k + a_k K(\theta_k, \tilde{S}_{k + 1}),
\end{equation}
where $\theta_k^T = ((\theta^{(1)}_k)^T,\dots,(\theta^{(n)}_k)^T)$ is the estimate of $\theta^*$ at time $k$, and $a_k$ is the step size.

\begin{rmk}\label{rmk1}
In this algorithm, we assume that $\theta_k$ is bounded. 
If this assumption does not hold, one can apply stochastic approximation algorithms with expanding truncations (see Appendix \ref{AppendixD} or \cite{chen2002stochastic}).
{It is verified that the number of truncations is finite a.s., so estimate $\theta_k$ is also bounded because of truncations.}
\end{rmk} 

\subsection{Asymptotic Properties}

In this {subsection} we provide the results on asymptotic properties of the proposed algorithm, including strong consistency and convergence rate. First, we introduce the following step size condition, which is standard for SA algorithms.
\begin{asmp}\label{at}
Let $a_k$ be the step size in \eqref{IdenAlg}, satisfying $a_k > 0$, $\sum\nolimits_{k = 1}^{\infty} a_k = \infty$, and $\sum\nolimits_{k = 1}^{\infty} a_k^2 < \infty$.
\end{asmp}

Under Assumptions \ref{noiseAsmpP} and \ref{at}, we have the following strong consistency result, indicating that Algorithm \eqref{IdenAlg} converges to the true parameter vector $\theta^*$.

\begin{thm}(Strong consistency)\label{mainthm2}\\
Suppose that Assumptions \ref{noiseAsmpP} and \ref{at} hold, then estimates $\theta_k$ of Algorithm \eqref{IdenAlg} converge to $\theta^*$ a.s., that is,
\[P\{\lim_{k\to\infty} \theta_k = \theta^*\} = 1,\]
from any fixed initial value, where $\theta^*$ is the true parameter vector.
\end{thm}

\begin{proof}
See Appendix \ref{AppendixD}.
\end{proof}

\begin{rmk}
Theorem \ref{mainthm2} establishes a theoretical guarantee for Algorithm \eqref{IdenAlg}, showing that the estimation problem of the weighted adjacency matrix can be solved under the independent standard Gaussian assumption.
\end{rmk}

For convergence rate, we prove that by choosing an appropriate step size, our proposed algorithm can have a convergence rate arbitrarily close to $O(1/\sqrt{k})$ a.s. Three hyper-parameters are given in the step size, which can be tuned to promote the performance of the algorithm in practice.

\begingroup
\def\theasmp{\ref{at}$^\prime$}
\addtocounter{asmp}{-1}
\begin{asmp}\label{atP}
Let $a_k$ be the step size in \eqref{IdenAlg}, satisfying $a_k = \frac{a}{k^{1 - \beta} + \gamma}$ with $a, \gamma > 0$ and $\beta \in (0, 1/3)$.
\end{asmp}
\endgroup

\begin{thm}\label{ConvergenceRateThm}(Convergence rate)
Suppose that Assumption \ref{noiseAsmpP} holds. If Assumption \ref{atP} holds, then for $\theta_k$ in Algorithm \eqref{IdenAlg},
\begin{equation}\label{deltaoriginal}
\|\theta_k - \theta^*\| = o(k^{-\delta}), ~\forall \delta \in (0, \frac12 - \beta), ~\text{a.s.}
\end{equation}
If Assumption \ref{atP} holds but with $\beta = 0$, then there exists $\delta' \in (0, 1/2]$ such that
\begin{equation}\label{deltaprime}
\|\theta_k - \theta^*\| = o(k^{-\delta}), ~\forall \delta \in (0, \delta'), ~\text{a.s.,}
\end{equation}
where $\delta'$ depends on the true parameter vector $\theta^*$.
\end{thm}

\begin{proof}
See Appendix \ref{AppendixE}.
\end{proof}

\begin{rmk}
Theorem \ref{ConvergenceRateThm} further characterizes the performance of \eqref{IdenAlg}, whose convergence rate can be arbitrarily close to the fastest rate of SA algorithms, i.e. $O(1/\sqrt{k})$, \cite{chen2002stochastic}.
In Theorem \ref{ConvergenceRateThm}, when the step size is selected as the order $O(1/k)$, the convergence rate may slow down, similar to a result on estimation for unknown thresholds of quantized output systems \cite{song2018recursive}. 
The bound $\delta'$ is related to the parameters to be estimated (and may not be a sharp bound), which is demonstrated in Section \ref{consistencySimulation}.
Since $a_k$ in Assumption \ref{atP} is a slowly decreasing step size \cite{chen2002stochastic}, applying the averaging technique may lead to the asymptotic efficiency of the algorithm, which will be a future work.
\end{rmk}

\section{Numerical Simulations}\label{numerical simulations}

In this section, we first demonstrate the asymptotic properties of the proposed algorithm, then apply it to an estimation problem of a Boolean network, and finally investigate the sensitivity of the algorithm under three unmodeled factors.

\subsection{Consistency and Convergence Rate}\label{consistencySimulation}

This subsection illustrates asymptotic properties of Algorithm \eqref{IdenAlg}. We set $n=2$ and randomly generate $A$ and $\bm{c}$ as
\begin{align*}
A = \begin{pmatrix} 0.87 & 0.13 \\ 0.62 & 0.38 \end{pmatrix}, \quad \bm{c} = (0.5~0.1)^T.
\end{align*}
The step size is set to be $a_k = 15 / (k + 100)$, and the algorithm is run for $200$ trials. Fig. \ref{consistencyFigure} shows the strong consistency of the algorithm. 
In both sub-figures, blue lines represent one sample path, and red lines represent the true value. 
Gray areas illustrate error bands for all $200$ trials.

Illustrations of the convergence rate of the proposed algorithm are shown in Fig. \ref{convergenceRateFig}, where the mean square errors (MSE) {for different parameter settings are presented. The MSE is defined as $\text{MSE}_k := \frac1N \sum_{i = 1}^N \|\theta_k^{[i]} - \theta^*\|^2$ with $N=200$ and $\theta_k^{[i]}$ being the $i$-th trial's estimate at time $k$.} 
According to Theorem \ref{ConvergenceRateThm}, the fastest speed of MSE is approximately of order $O(1/k)$, and the proposed algorithm with step size $a_k = 15 / (k + 100)$ ($\beta = 0$ for $a_k$ in Assumption \ref{atP}) can actually achieve this speed for $c_1 = 0.5$. 
But the convergence rate under this step size may slow down as the true parameter vector becoming ``worse'', as shown in Fig. \ref{rate1/2}, where simulations are conducted with $c_1$ to be $0.5$, $1$, $1.5$, and $2$, and the step size $a_k = 15 / (k + 100)$. 
This illustrates \eqref{deltaprime} in Theorem \ref{ConvergenceRateThm}. 

For $c_1 = 2$ and step size $a_k = 15 / (k^{1-\beta} + 100)$ with $\beta = 0$, $0.01$, $0.05$, and $0.1$, simulation results are presented in Fig. \ref{rate2}. 
The convergence rate of the algorithm seems to slow down as $\beta$ decreasing to zero, but this does not contradict with \eqref{deltaoriginal} of Theorem \ref{ConvergenceRateThm}. This is because the latter demonstrates an asymptotic behavior, and it can be observed that the decreasing speed of MSEs under nonzero $\beta$ is getting larger as $k$ increasing in Fig. \ref{rate2}. Moreover, as in the proof of Theorem \ref{ConvergenceRateThm}, the upper bound of $\delta$ depends on the limit of $a_{k+1}^{-1} - a_{k}^{-1}$. For $a_k = \frac1{k^{1-\beta}}$, $a_{k+1}^{-1} - a_{k}^{-1} = O(\frac{1}{k^{\beta}})$, indicating that if $\beta$ close to zero, then $a_{k+1}^{-1} - a_{k}^{-1}$ has an extremely slow decreasing speed. This may result in a similar transient behavior for the proposed algorithm with $\beta$ close to zero and $\beta = 0$.

The above simulations show that the best choice of $\beta$ for $a_k = \frac{a}{k^{1 - \beta} + \gamma}$ may depend on data size and unknown parameters. If data size is large enough, one can set $\beta$ to be or close to zero, in order to achieve a fast convergence rate. But if data size is small and the unknown parameters are not ``good enough'', for example, the system providing less diverse information, then one may obtain a better estimation result with a relatively large $\beta$.

\begin{figure}[t]
\centering
\subfigure[\label{estimationA}The estimation of $A$.]{
\includegraphics[scale=0.34]{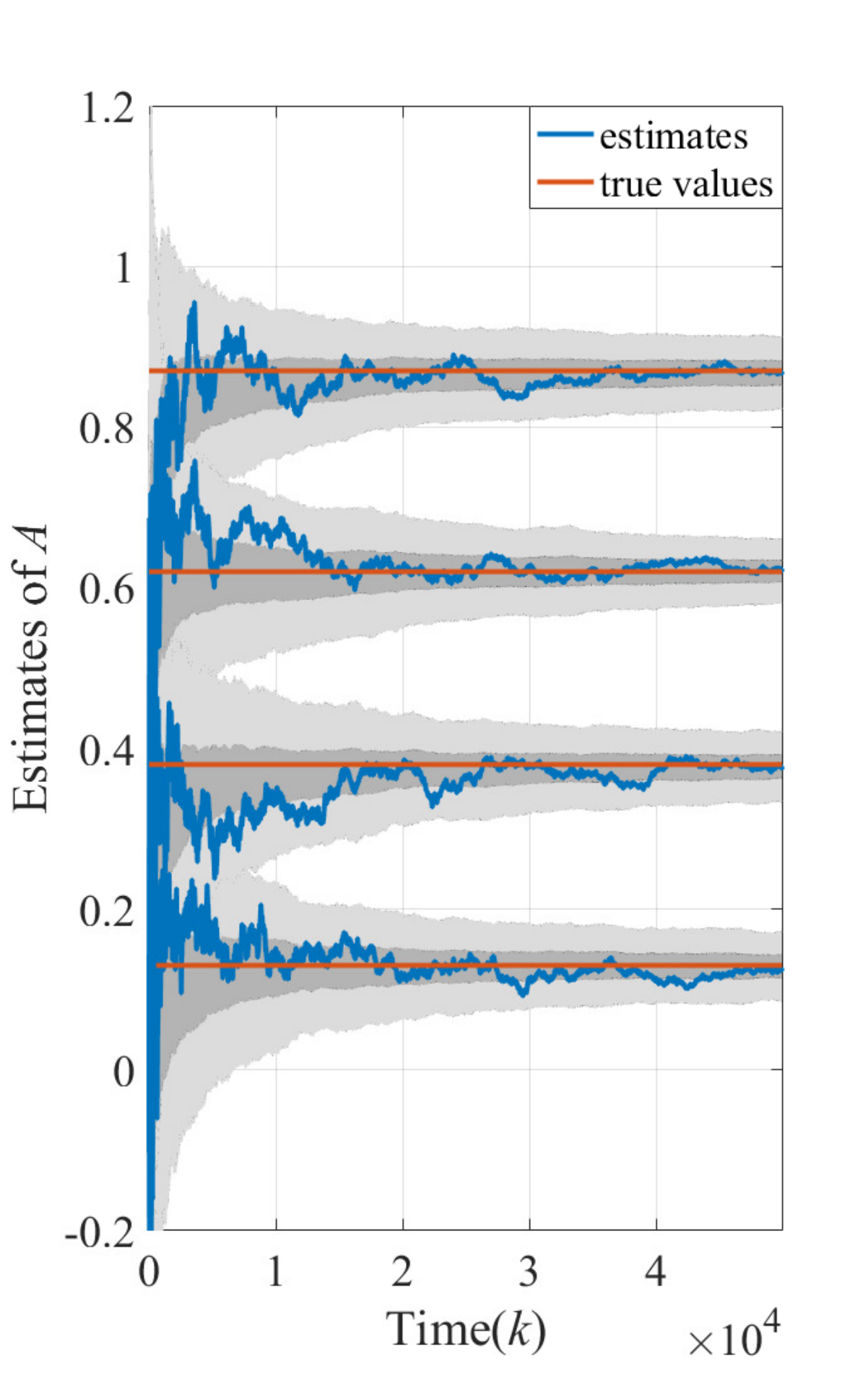}}
\subfigure[\label{estimationC}The estimation of $\bm{c}$.]{
\includegraphics[scale=0.34]{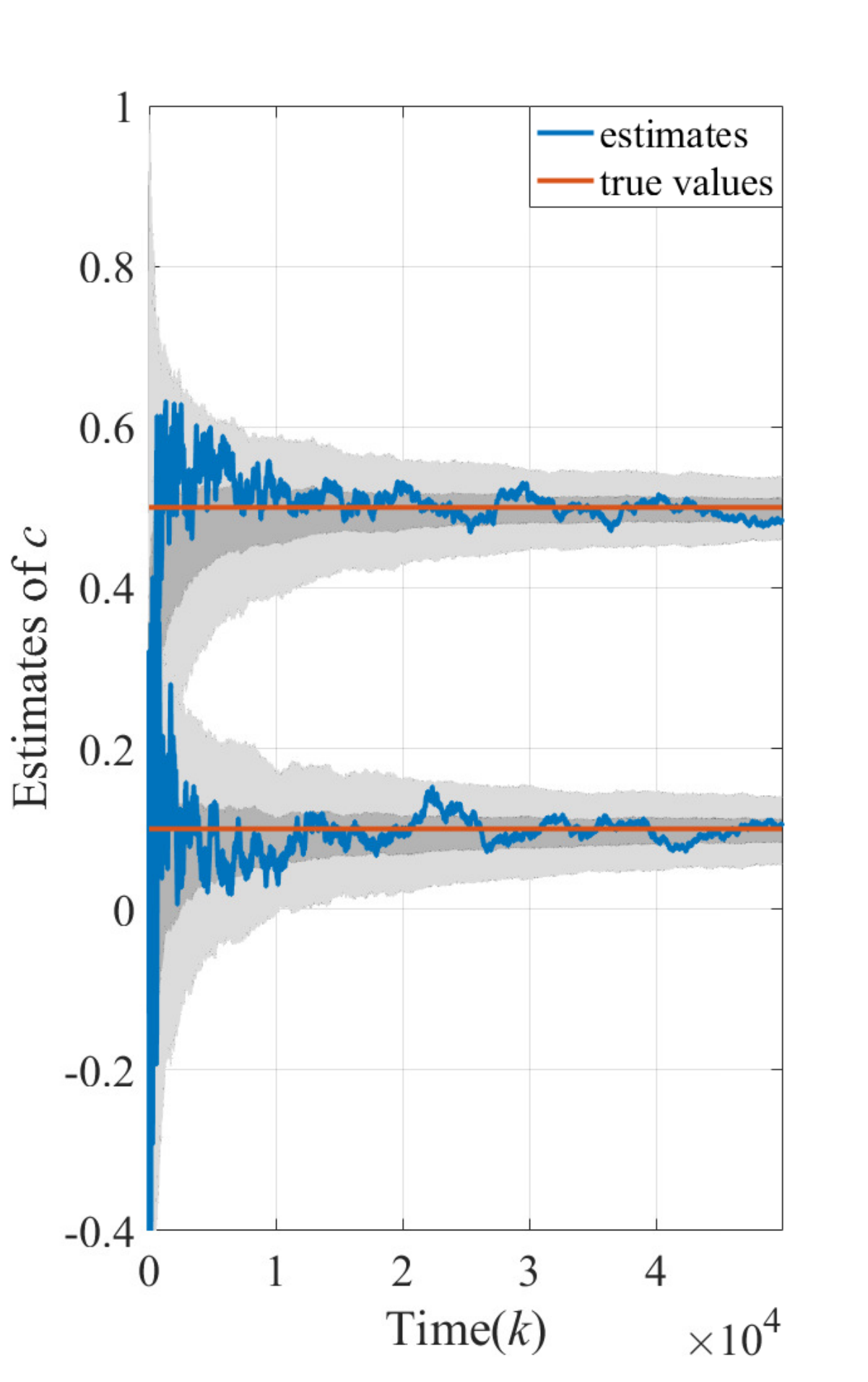}}
\caption{\label{consistencyFigure}Consistency of Algorithm \eqref{IdenAlg}. The dark (light) gray areas demonstrate error bands with one (three) standard deviation.}
\end{figure}
\begin{figure}[t]
\centering
\subfigure[\label{rate1/2}Cases of $\beta = 0$ and different $c_1$.]{
\includegraphics[scale=0.4]{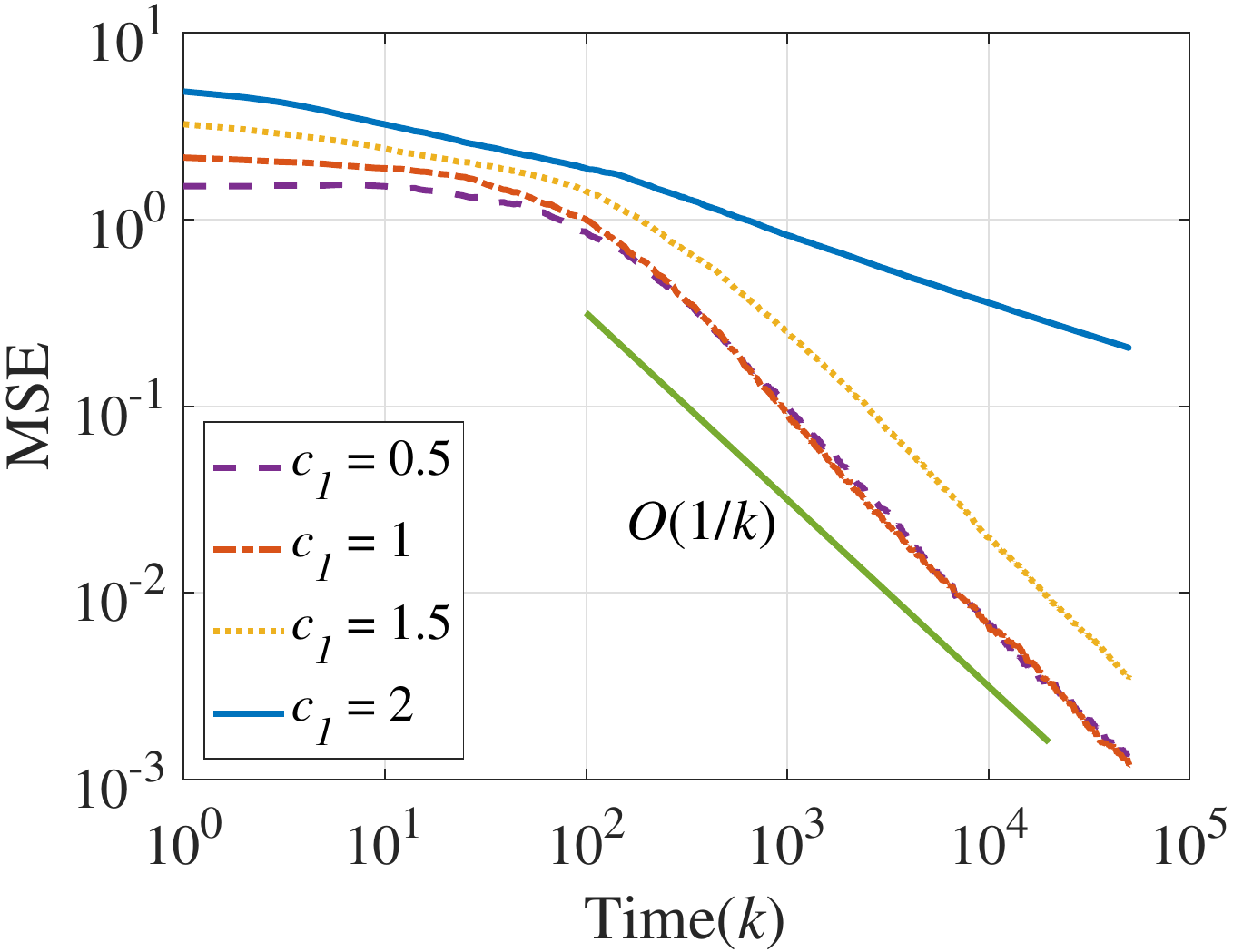}}
\subfigure[\label{rate2}Cases of $c_1 = 2$ and different $\beta$.]{
\includegraphics[scale=0.4]{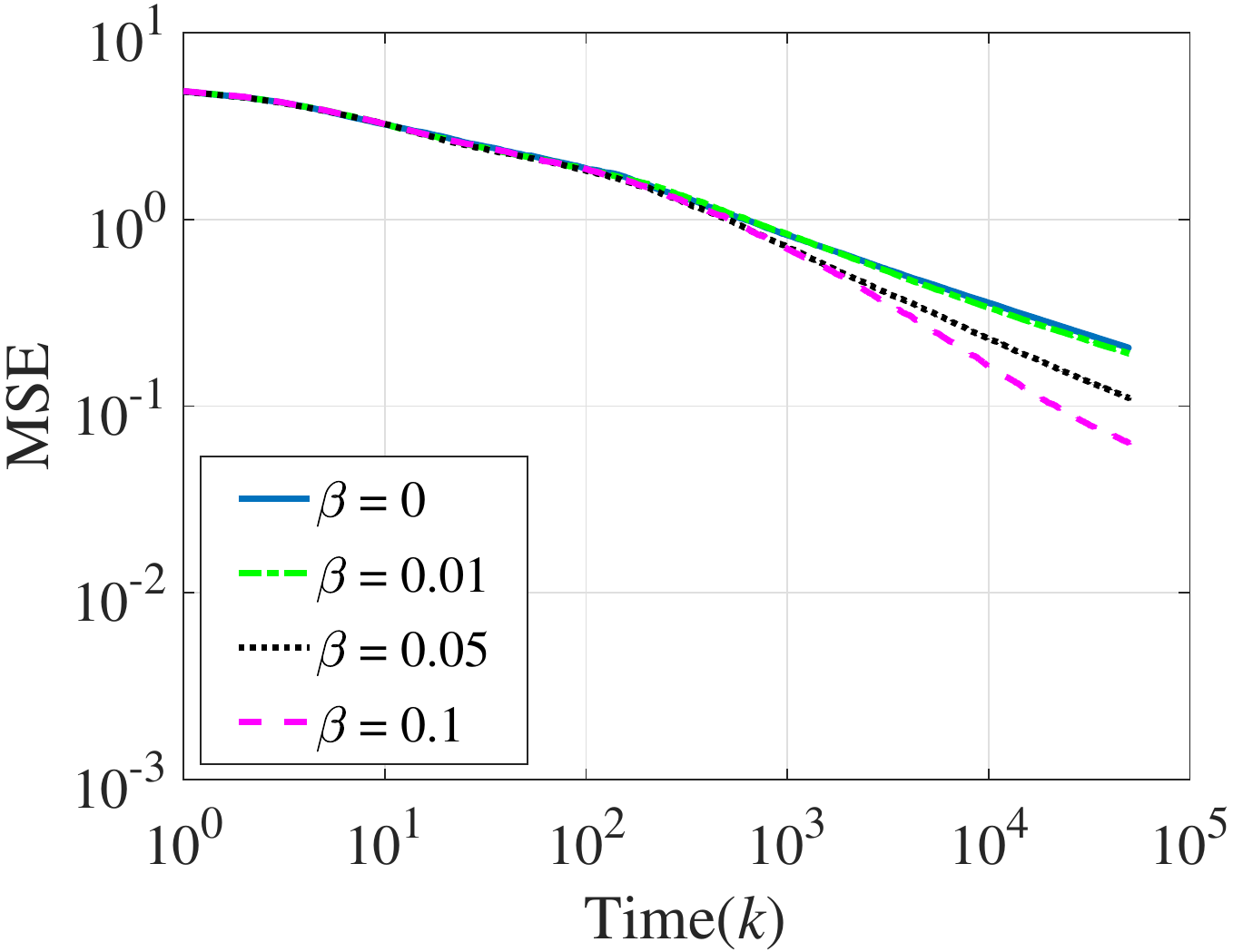}}
\caption{\label{convergenceRateFig}Convergence rate of Algorithm \eqref{IdenAlg}.}
\end{figure}

\begin{figure}[t]
\centering
\subfigure[\label{yeast_true}The true yeast cell-cycle network.]{
\includegraphics[scale=0.4]{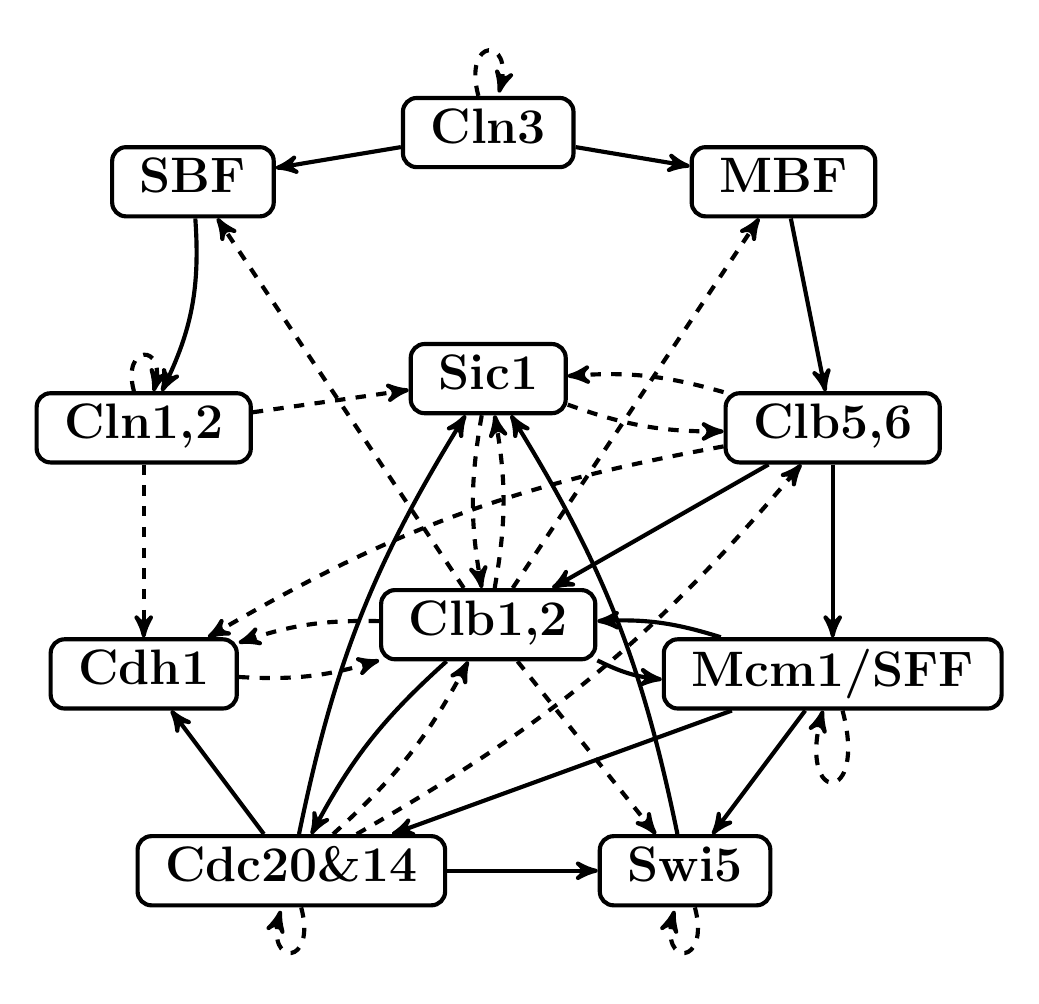}}
\subfigure[\label{yeast_esti}The estimated yeast network.]{
\includegraphics[scale=0.4]{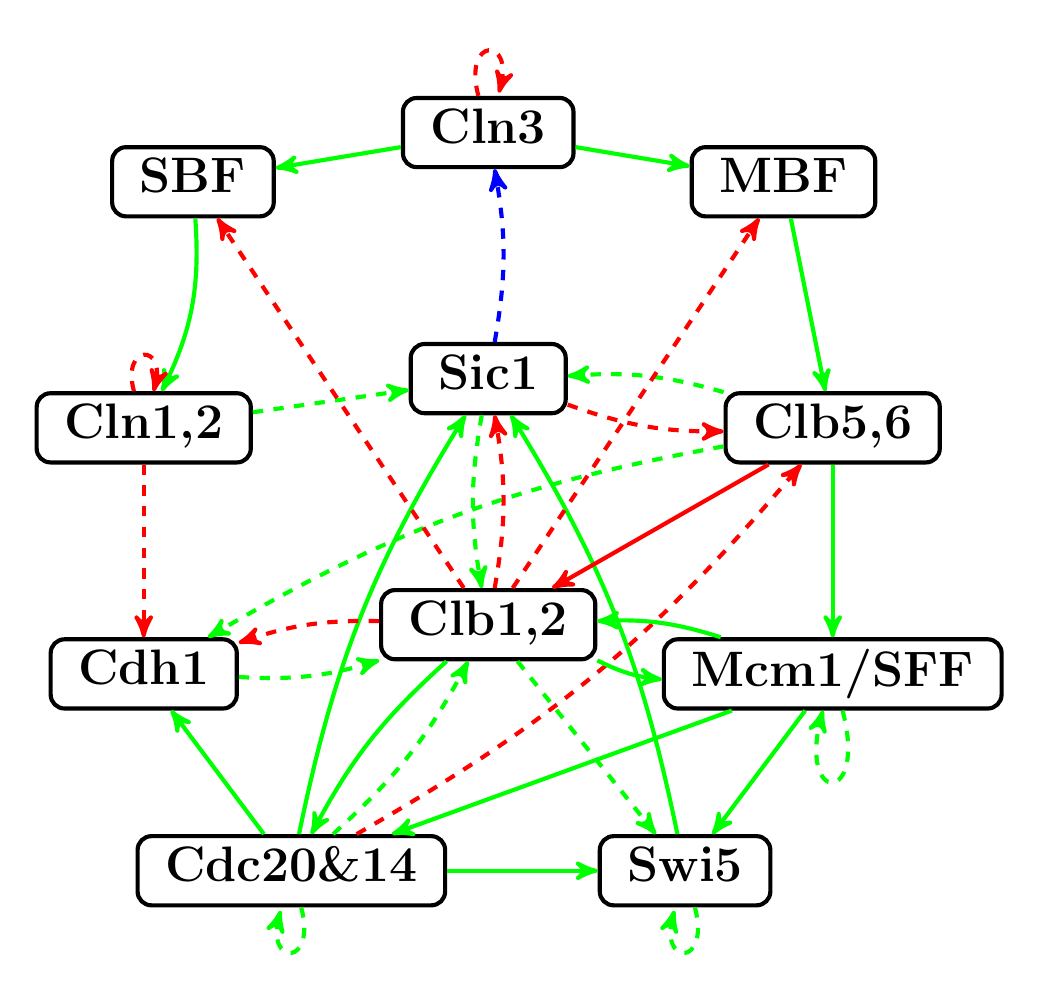}}
\caption{Numerical experiment for estimation of a yeast cell-cycle network \cite{li2004yeast}. In both sub-figures, arrows indicate activation while dashed arrows indicate inhibition. In (b), green means that an edge is correctly inferred, red means that an edge exists but is not detected by Algorithm \eqref{IdenAlg}, and blue means false detection.}
\end{figure}

\begin{figure*}[t]
\centering
\subfigure[\label{asyn}The asynchronous update scenario.]{
\includegraphics[scale=0.4]{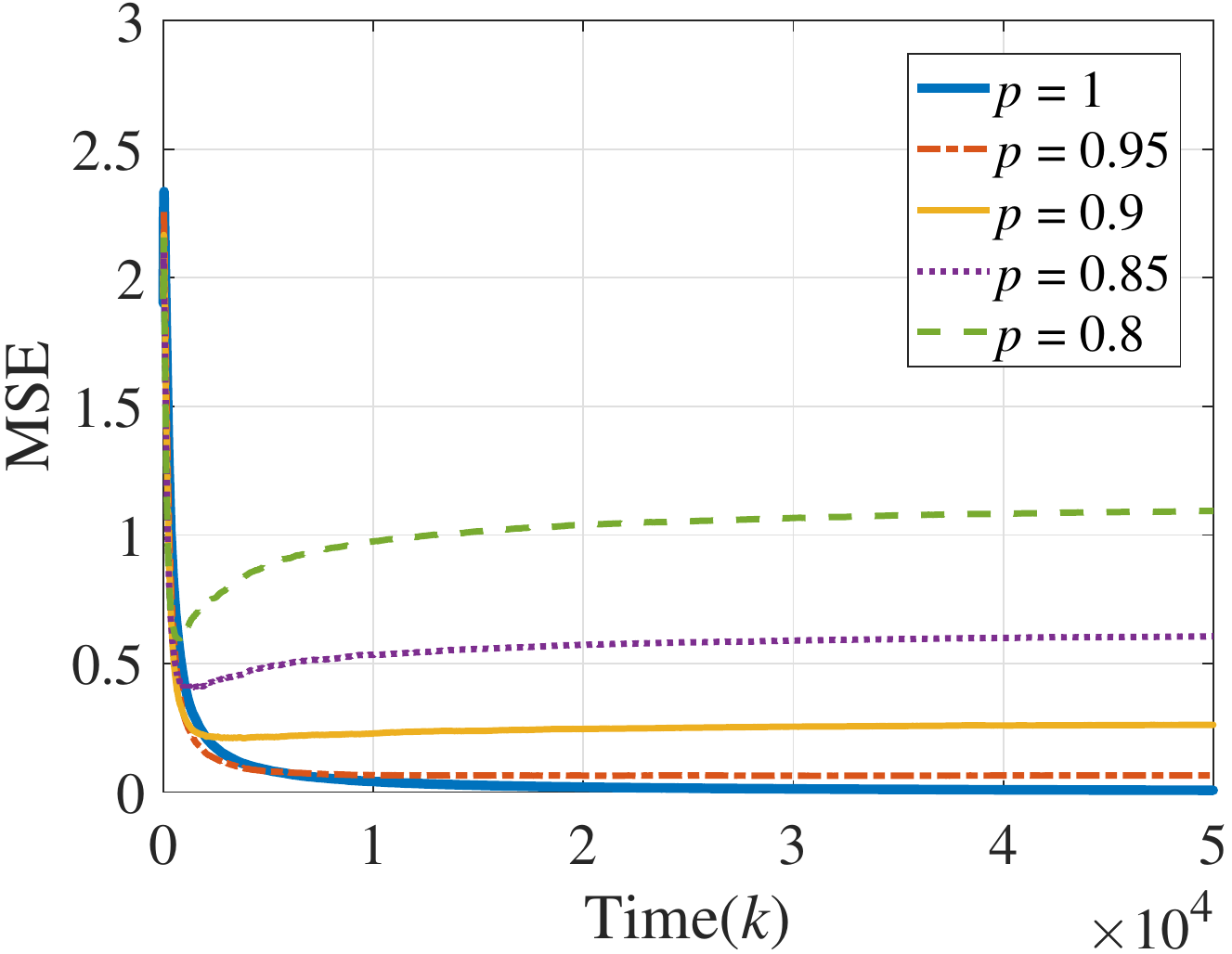}}
\subfigure[\label{inertia}The occasional disturbance scenario.]{
\includegraphics[scale=0.4]{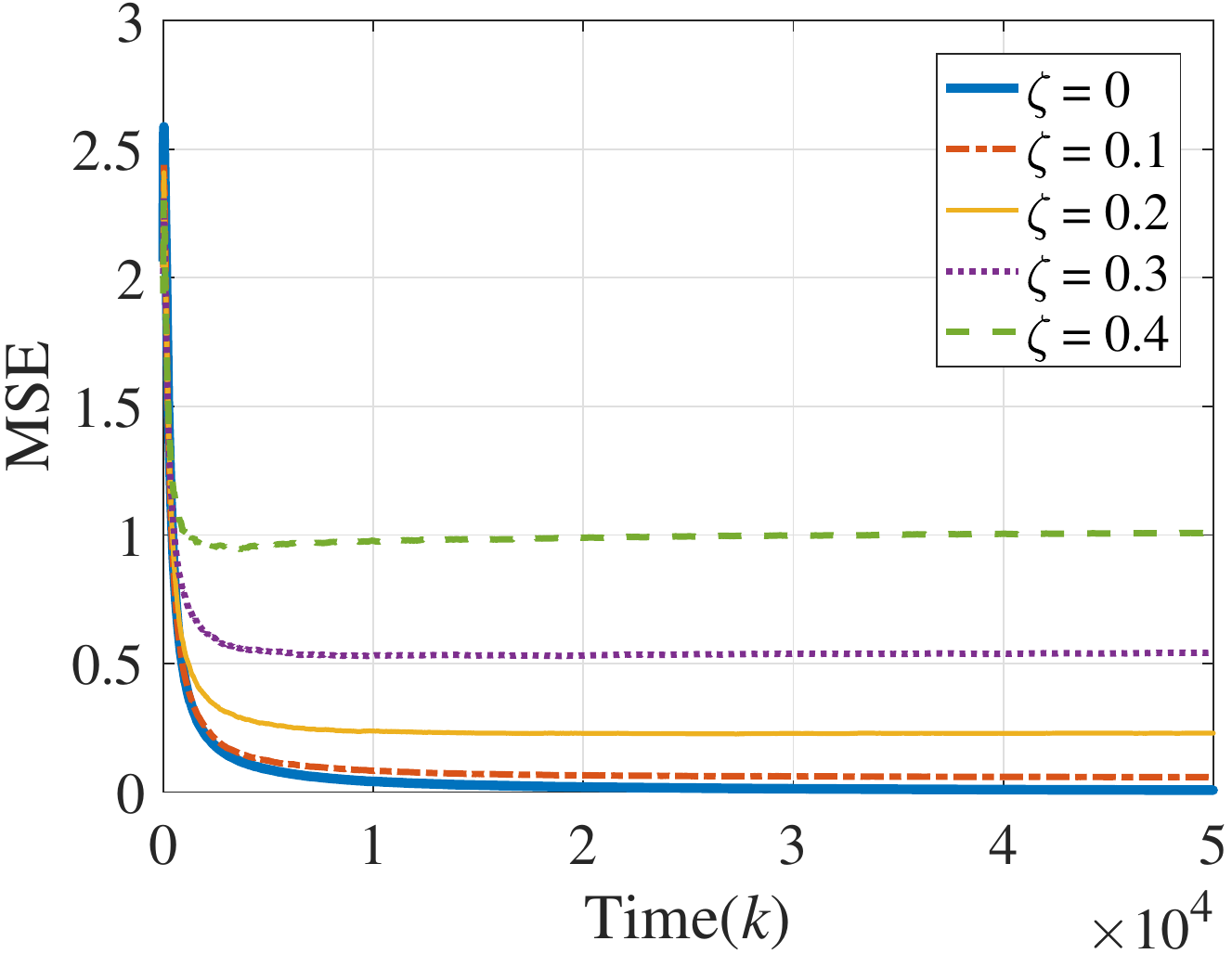}}
\subfigure[\label{linkf}The time-varying communication scenario.]{
\includegraphics[scale=0.4]{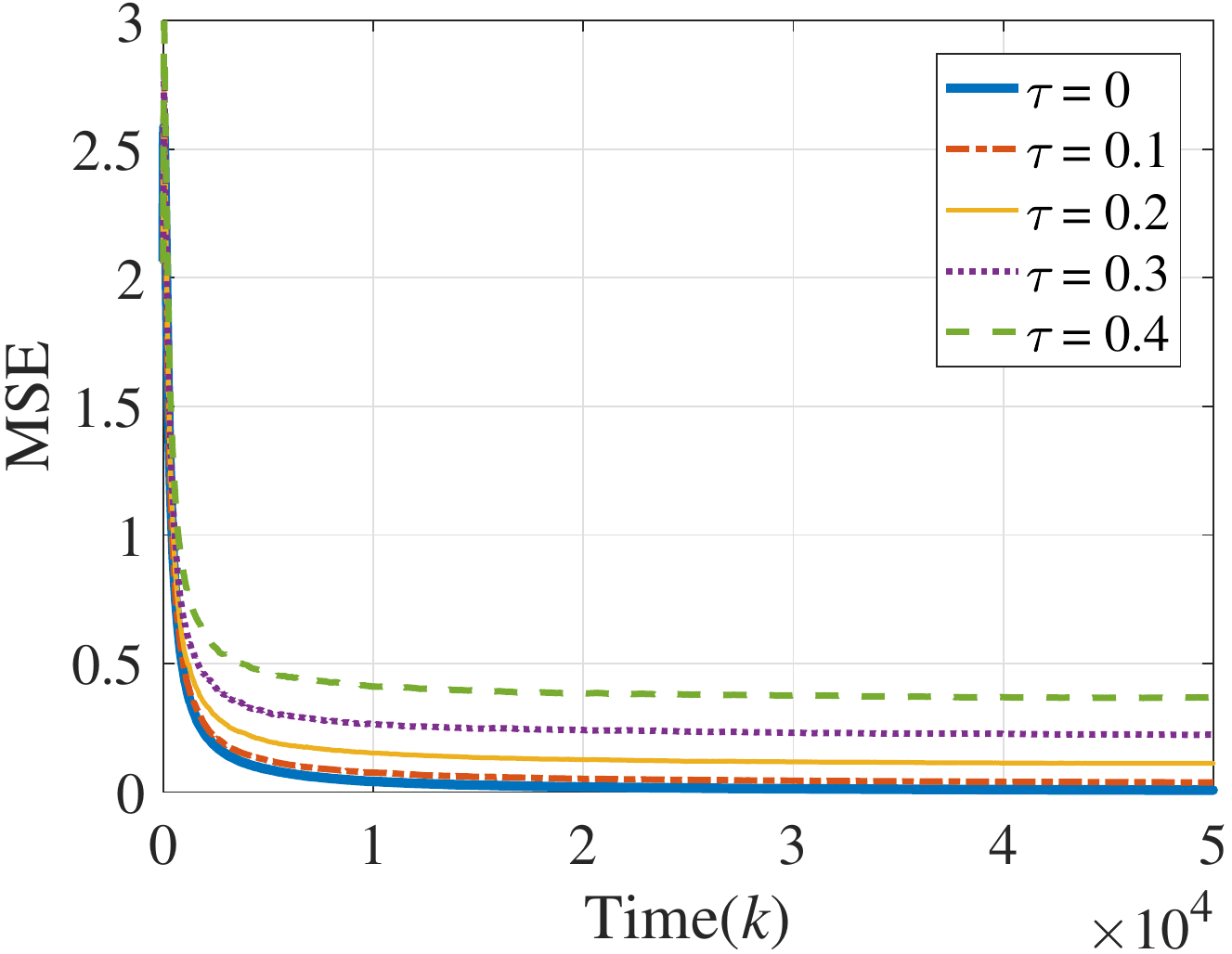}}
\caption{MSE illustrations for sensitivity investigation in Section \ref{sensitivitySimulation}.}
\end{figure*}

\subsection{Numerical Experiment}\label{experiment}

In this section we apply our algorithm to estimate a Boolean network with perturbation. In the numerical experiment, a yeast cell-cycle network \cite{li2004yeast} is used, shown in Fig. \ref{yeast_true}, and a sequence of observation data of length $500$ is generated according to \eqref{exampleAEquation} in Example $1$ with $\eta = 0.05$. 
In this system, weights of activation (or inhibition) relationships between different nodes are assumed to be $1$ (or $-1$), and self weight of an node $i$ is taken to be $-(d_i^{\text{in}}+1)$, representing self-degradation \cite{li2004yeast}, where $d_i^{\text{in}}$ is the in-degree of $i$. Although Theorem \ref{unident} indicates non-identifiability of the system, it could be identifiable for integer-valued weights \cite{akutsu2018algorithms}. 
Thus, we simply round the estimate of adjacency weights at the final step. 
The structural estimation result, i.e., estimation of the existence and sign of edges, is shown in Fig. \ref{yeast_esti}. 
The result is relatively good despite of small-sampled data, since most of the existing edges are detected and there is few false detection.

\subsection{Sensitivity}\label{sensitivitySimulation}

We now study the influence of three possible unmodeled factors on the performance of our algorithm. An influence weight matrix with four individuals from an empirical study \cite{friedkin1990social} is selected for investigation. The weighted adjacency matrix $A$ is given by
\begin{equation*}
A = \begin{pmatrix} 0.44 & 0.12 & 0.36 & 0.3 \\
0.147 & 0.215 & 0.344 & 0.294 \\
0 & 0 & 1 & 0 \\
0.09 & 0.178 & 0.446 & 0.286
\end{pmatrix}.
\end{equation*}
The disturbances are set to be independent standard Gaussian, and $\bm{c}$ is randomly generated as $(0.13~0.28~0.08~0.24)^T$.

The first disturbing factor of our concern is that agents may update their outputs asynchronously, which is a  common phenomenon and considered, for instance, in the original paper of Example $2$ \cite{blume2003equilibrium}. 
For simplicity, suppose that at each time agents update their outputs mutually independently and with probability $p \in (0,1]$. 
When $p=1$, the system becomes \eqref{binary}.  For different updating probability $p$, we compute the MSE with the number of trials $N = 200$. 
The result is shown in Fig. \ref{asyn}, and it can be found that the proposed algorithm only performs well for $p$ close to one. 
This indicates that estimating adjacency matrix without update information is a relatively tricky task, since one cannot know when an update happens with the presence of random disturbance. 
But if updates are known, then our algorithm can tackle this issue. This is because Assumption \ref{noiseAsmpP} decouples the estimation problem for different agents.

The second factor is that the disturbance occurs less frequently; more specifically, for $\zeta \in [0, 1)$ and $k \ge 0$, $1 \le i \le n$, set 
\begin{equation}\label{noise2}
D_{k, i}' = 
\begin{cases}
0 & \text{ with probability } \zeta\\
D_{k, i} & \text{ with probability } 1 - \zeta,
\end{cases}
\end{equation}
and the true system is as follows
\begin{equation}\label{binary2}
\begin{aligned}
Y_{k+1}&= A S_k + D_k',\\
S_k &= \mathcal{Q}(Y_k, \bm{c}).
\end{aligned}
\end{equation}
In other words, the agents are only affected by disturbances occasionally. When $\zeta$ is large, the disturbance behaves more like a step or pulse signal \cite{ljung1987system}. The MSEs with $N = 200$ for different $\zeta$ in this case are illustrated in Fig. \ref{inertia}. If $\zeta = 0$, then \eqref{binary2} is the same as \eqref{binary}, and the MSE converges to zero because of strong consistency. As $\zeta$ grows larger, the error of the algorithm increases, but the MSE remains small for $\zeta$ around $0.1$.

The final scenario considered here is that the network is time-varying because of environment randomness or communication outages. 
Let $\{u_{k, ij}\}$, $k \ge 0$, $1 \le i, j \le n$, be a sequence of i.i.d. Bernoulli random variables, taking value $0$ with probability $\tau$ and value $1$ with probability $1 - \tau$. The true dynamic for $Y_{k}$ is as below
\begin{equation*}
Y_{k+1, i} = \sum_{j \in \mathcal{V}} a_{ij} u_{k, ij} S_{k, j} + D_{k,i},
\end{equation*}
which means that agent $i$ does not receive the output of $j$ with probability $\tau$. Similarly, the MSEs are demonstrated in Fig. \ref{linkf}. 
The case of $\tau = 0$ represents the original system, so the algorithm converges in the end. The estimation error becomes greater for larger $\tau$, but the algorithm performs better than the two former cases.
To sum up, our algorithm is insensitive to small unmodeled factors.

\section{Conclusion}\label{conclusion}
In this paper we studied a recursive adjacency matrix estimation problem based on binary data.
Stability of the observation sequence and identifiability of the model {parameters} were studied. 
We followed a maximum likelihood approach to address the network estimation issue, and developed a recursive algorithm based on SA techniques to solve the estimation problem. 
The strong consistency of the algorithm was verified, and its convergence rate analyzed. 
Future work includes investigation of asymptotical efficiency of the algorithm, as well as generalization of the model and noise conditions.

\appendices

\section{}\label{AppendixA}

\noindent\emph{\textbf{Proof of {Lemma} \ref{lem1}}:} Let $\tilde{P}$ be the transition probability matrix of $\{\tilde{S}_k\}$. From the definition of stationary distribution, we have that
\begin{equation*}
P\{\tilde{S} = \tilde{\bm{s}}\} = \sum_{\tilde{\bm{u}} \in \mathcal{S}^{2n}} P\{\tilde{S} = \tilde{\bm{u}}\} \tilde{P} (\tilde{\bm{u}}, \tilde{\bm{s}}).
\end{equation*}
Define $\mathscr{S}_F := \{\tilde{\bm{u}} \in \mathcal{S}^{2n} : \tilde{\bm{u}}_{1:n} = \tilde{\bm{s}}_{n+1:2n}\}$, and it follows from the definition of $\{\tilde{S}_k\}$ that $P(\tilde{\bm{u}}, \tilde{\bm{s}}) = 0$ for $\tilde{\bm{u}} \not \in \mathscr{S}_F$. Hence, 
\begin{equation}\label{lem1eq1}
P\{\tilde{S} = \tilde{\bm{s}}\} = \sum_{\tilde{\bm{u}} \in \mathscr{S}_F} P\{\tilde{S} = \tilde{\bm{u}}\} \tilde{P} (\tilde{\bm{u}}, \tilde{\bm{s}}).
\end{equation}
Similarly, we have that
\begin{align}\nonumber
P\{\tilde{S}_{n+1:2n} = \tilde{\bm{s}}_{n+1:2n}\} &= \sum_{{\tilde{\bm{v}}} \in \mathscr{S}_L} P\{\tilde{S} = {\tilde{\bm{v}}}\}\\\label{lem1eq2}
&= \sum_{\tilde{\bm{u}} \in \mathscr{S}_F} \sum_{\tilde{\bm{v}} \in \mathscr{S}_L} P\{\tilde{S} = \tilde{\bm{u}}\} \tilde{P} (\tilde{\bm{u}}, \tilde{\bm{v}}),
\end{align}
where $\mathscr{S}_L := \{\tilde{\bm{v}} \in \mathcal{S}^{2n} : \tilde{\bm{v}}_{n+1:2n} = \tilde{\bm{s}}_{n+1:2n}\}$. Combining \eqref{tildeS} with \eqref{lem1eq1} and \eqref{lem1eq2} respectively, {it} holds that
\begin{align*}
&P\{\tilde{S} = \tilde{\bm{s}}\} = \sum_{\tilde{\bm{u}} \in \mathscr{S}_F} P\{\tilde{S} = \tilde{\bm{u}}\} P (\tilde{\bm{s}}_{n+1:2n}, \tilde{\bm{s}}_{1:n}),\\
&P\{\tilde{S}_{n+1:2n} = \tilde{\bm{s}}_{n+1:2n}\} \\
&= \sum_{\tilde{\bm{u}} \in \mathscr{S}_F} \sum_{\tilde{\bm{v}} \in \mathscr{S}_L} P\{\tilde{S} = \tilde{\bm{u}}\} P (\tilde{\bm{s}}_{n+1:2n}, \tilde{\bm{v}}_{1:n})\\
&= \sum_{\tilde{\bm{u}} \in \mathscr{S}_F} \sum_{\bm{w} \in \mathcal{S}^n} P\{\tilde{S} = \tilde{\bm{u}}\} P (\tilde{\bm{s}}_{n+1:2n}, \bm{w})\\
&= \sum_{\tilde{\bm{u}} \in \mathscr{S}_F} P\{\tilde{S} = \tilde{\bm{u}}\}.
\end{align*}
Hence, 
\begin{align*}
&P\{\tilde{S}_{1:n} = \tilde{\bm{s}}_{1:n} | \tilde{S}_{n+1:2n} = \tilde{\bm{s}}_{n+1:2n}\} \\
&=\frac{P\{\tilde{S} = \tilde{\bm{s}}\}}{P\{\tilde{S}_{n+1:2n} = \tilde{\bm{s}}_{n+1:2n}\}}\\
&=\frac{\sum_{\tilde{\bm{u}} \in \mathscr{S}_F} P\{\tilde{S} = \tilde{\bm{u}}\} P (\tilde{\bm{s}}_{n+1:2n}, \tilde{\bm{s}}_{1:n})}{\sum_{\tilde{\bm{u}} \in \mathscr{S}_F} P\{\tilde{S} = \tilde{\bm{u}}\}} = P (\tilde{\bm{s}}_{n+1:2n}, \tilde{\bm{s}}_{1:n}).
\end{align*}
\hfill$\Box$

\begingroup
\def\thethm{\ref{MC}$^\prime$}
\addtocounter{thm}{-1}
\begin{thm}\label{MCprime}
Suppose that Assumption \ref{noiseAsmpP} holds, then Markov chain $\{S_k\}$ is irreducible and aperiodic. Hence, it converges in distribution, from any initial condition, to a unique stationary distribution $\pi$ on $\mathcal{S}^n$ with $\pi(\bm{s}) > 0$, $\forall \bm{s} \in \mathcal{S}^n$.
\end{thm}
\endgroup

\begingroup
\def\thethm{\ref{MC2}$^\prime$}
\begin{thm}\label{MC2prime}
Suppose that Assumption \ref{noiseAsmpP} holds, then Markov chain $\{\tilde{S}_k\}$ is irreducible and aperiodic. Hence, it converges in distribution, from any initial condition, to a unique stationary distribution $\tilde{\pi}$ on $\mathcal{S}^{2n}$ with $\tilde{\pi}(\tilde{\bm{s}}) > 0$, $\forall \tilde{\bm{s}} \in \mathcal{S}^{2n}$.
\end{thm}
\endgroup

\begingroup
\def\thelem{\ref{lem1}$^\prime$}
\addtocounter{lem}{-1}
\begin{lem}\label{lem1prime}
Suppose that Assumption \ref{noiseAsmpP} holds, and $\tilde{S}$ is subject to the stationary distribution of $\{\tilde{S}_k\}$. Then 
\begin{equation*}
P\{\tilde{S}_{1:n} = \tilde{\bm{s}}_{1:n} | \tilde{S}_{n+1:2n} = \tilde{\bm{s}}_{n+1:2n}\} = P(\tilde{\bm{s}}_{n+1:2n}, \tilde{\bm{s}}_{1:n}),
\end{equation*}
$\forall \tilde{\bm{s}} \in \mathcal{S}^{2n}$, where $P(\cdot, \cdot)$ is the probability transition matrix of $\{S_k\}$.
\end{lem}
\endgroup

\section{}\label{AppendixB}
\noindent\emph{\textbf{Proof of {Theorem} \ref{unident}}:}
Without loss of generality, suppose that $a_{ij} \not= 0$ {from the assumption for $A$}. 
For $A_i$ with $A_i \bm{1}_n \not= 0$, either $0 < \frac12 \sum_{j=1}^n a_{ij} < \sum_{j=1}^n a_{ij}$ or $0 > \frac12 \sum_{j=1}^n a_{ij} > \sum_{j=1}^n a_{ij}$ holds, so from ii) of Assumption \ref{noiseAsmpPPP} {it} holds that $d_{i3} \ge (c_i - A_i \bm{0}_n) {\vee} (c_i - A_i \bm{1}_n) > c_i - \frac12 A_i \bm{1}_n = d_{i2} > (c_i - A_i \bm{0}_n) {\wedge} (c_i - A_i \bm{1}_n) > d_{i1}$. 
If $A_i \bm{1}_n = 0$, then there exists $a_{ik} < 0$.
Thus, $d_{i3} \ge (c_i - A_i \bm{e}_k) > c_i - \frac12 A_i \bm{1}_n = d_{i2} = (c_i - A_i \bm{0}_n) > d_{i1}$. Therefore, we have $d_{i3} > d_{i2} > d_{i1}$.

Hence, under Assumption \ref{noiseAsmpPPP} we have the following properties for $D_{k,i}$, $1\le i\le n$, $k\ge 0$:  

\noindent
$P\{D_{k,i} \ge c_i - A_i \bm{s}\} = P\{D_{k,i} = d_{i2} \text{ or } d_{i3}\} = (1 - 2 \eta) + \eta = 1 - \eta$ for $\bm{s} \in \mathcal{S}^n$ such that $A_i (\bm{s} - \frac12 \bm{1}_n) \ge 0$, and $P\{D_{k,i} \ge c_i - A_i \bm{s}\} = P\{D_{k,i} = d_{i3}\} = \eta$ for $\bm{s} \in \mathcal{S}^n$ such that $A_i(\bm{s} - \frac12 \bm{1}_n) < 0$.

Since $\mathcal{S}^n$ is a finite set, there exists $\hat{A}_i \not = A_i$ such that $d_{i1} < c_i - \hat{A}_i\bm{s} \le c_i - A_i \bm{s}$ (implying $\hat{A}_i\bm{s} \ge A_i \bm{s}$) for all $\bm{s} \in \mathcal{S}^n$, and $\hat{A}_i\bm{s} - \frac12 A_i \bm{1}_n < 0$ for $\bm{s}$ such that $A_i(\bm{s} - \frac12 \bm{1}_n) < 0$ (if such $\bm{s}$ exists). This can be done by adding a sufficiently small constant to one of the entries of $A_i$.

Hence, $d_{i2} - (c_i - \hat{A}_i \bm{s}) = \hat{A}_i \bm{s} - \frac12 A_i \bm{1}_n \ge A_i \bm{s} - \frac12 A_i \bm{1}_n\ge 0$ for those $\bm{s}$ such that $A_i (\bm{s} - \frac12 \bm{1}_n) \ge 0$, implying $P\{D_{k,i} \ge c_i - \hat{A}_i \bm{s}\} = P\{D_{k,i} = d_{i2} \text{ or } d_{i3}\} = 1 - \eta$ for these $\bm{s}$. 
On the other hand, for those $\bm{s}$ such that $A_i(\bm{s} - \frac12 \bm{1}_n) < 0$, $P\{D_{k,i} \ge c_i - \hat{A}_i \bm{s}\} = P\{D_{k,i} = d_{i3}\} = \eta$, because $d_{i2} - (c_i - \hat{A}_i \bm{s}) = \hat{A}_i\bm{s} - \frac12 A_i \bm{1}_n < 0$. 
Therefore, $\theta = \text{vec}\big\{(A~\bm{c})\big\}$ and $\hat{\theta} = \text{vec}\big\{(\hat{A}~\bm{c})\big\}$, where $\hat{A} := (A_1^T~\cdots~A_n^T)^T$, define the same transition probability for observations, even though $\{D_{k,i}\}$ is fixed. 
\hfill$\Box$

\section{Proof of Theorem \ref{mainthm1}}\label{AppendixC}

We first introduce the following two lemmas.

\begin{lem}[\cite{ferguson2017course}{, pp. 124}]\label{passinglemma}
Let $(\Omega, \mathscr{F}, P)$ be a probability space, and let $f(\cdot, \cdot) : S \times \Omega \to \mathbb{R}$, where $S$ is an open interval, be a function satisfying:\\
\noindent
(i) $E \{ f(\theta, \omega)\}$ exists, $\forall \theta \in S$;
\\\noindent
(ii) $\frac{\partial}{\partial \theta} f(\theta, \omega)$ exists and is continuous in $\theta$, $\forall \omega \in \Omega$;
\\\noindent
(iii) there exists an integrable nonnegative function $g$ such that $\big|\frac{\partial}{\partial \theta} f(\theta, \omega)\big| \le g(\omega)$, $\forall (\theta, \omega) \in S \times \Omega$. \\\noindent
Then \[\frac{d}{d\theta} \int_{\Omega} f(\theta, \omega) dP(\omega) = \int_{\Omega} \frac{\partial}{\partial\theta} f(\theta, \omega) dP(\omega).\]
\end{lem}

\begin{lem}\label{G}~\\
$G(x) := \dfrac{x\phi(x)\Phi(x) + \phi^2(x)}{\Phi^2(x)} \in (0, C)$, for $x \in \mathbb{R}$, where $C$ is a positive constant, and $\phi(x)$ represents the probability density function of the standard Gaussian distribution.
\end{lem}

\begin{proof}
For $x \ge 0$, $G(x) > 0$ by definition. For $x < 0$, from the inequality (Lemma 2.3.3 in \cite{chow2012probability})
\begin{align*}
\Phi(x) < -\frac1x \phi(x), \quad x < 0,
\end{align*}
where $\Phi(x)$ represents the cumulative density function of the standard Gaussian distribution, {it} holds that $x\phi(x)\Phi(x) + \phi^2(x) > -\phi^2(x) + \phi^2(x) = 0$, and hence $G(x) > 0$. 
To prove that $G(x)$ has an upper bound, it suffices to note that {$\lim_{x \to -\infty}G(x) = 1$, which follows from the L'H{\^o}pital's rule and $\phi'(x) = -x\phi(x)$, and $\lim_{x \to +\infty} G(x) = 0$.}
\end{proof}

\noindent\emph{\textbf{Proof of Theorem \ref{mainthm1}}:} We divide the proof into four steps. 

\noindent\emph{Step 1.} For $\|\theta^{(i)}\| < M$, where $\theta^{(i)} = (A_i~c_i)^T$ and $M$ is an arbitrary positive real number, we use Lemma \ref{passinglemma} to show that the derivative can be passed under the expectation for functions $E\{\log g_i(\tilde{S}|\theta^{(i)})\}$ with $g_i$ defined in \eqref{g_i}, $1 \le i \le n$. That is to say, for $1 \le i \le n$ and $\|\theta^{(i)}\| < M$,
\begin{align}\label{passing1}
&\nabla_{\theta^{(i)}} E \{\log g_i(\tilde{S}|\theta^{(i)})\} = E \{\nabla_{\theta^{(i)}} \log g_i(\tilde{S}|\theta^{(i)})\},\\\label{passing2}
&\nabla^2_{\theta^{(i)}} E \{\log g_i(\tilde{S}|\theta^{(i)})\} = E \{\nabla^2_{\theta^{(i)}} \log g_i(\tilde{S}|\theta^{(i)})\}.
\end{align}
From the definition of $g_i$ in \eqref{g_i} and Assumption \ref{noiseAsmpP}, it follows that
\begin{align*}
\log g_i(\tilde{S}|\theta^{(i)})&= \log \big[(1 - \Phi(c_i - A_{i} S))^{\tilde{S}_i} \Phi(c_i - A_{i} S)^{1 - \tilde{S}_i} \big]\\
&\in [M_1, M_2],
\end{align*}
where $S := \tilde{S}_{n+1:2n}$ hereafter, $M_1 = \log (1 - \Phi(c_i + |A_i| \bm{1}_n)) \wedge \log \Phi(c_i - |A_i| \bm{1}_n)$, $M_2 = \log (1 - \Phi(c_i - |A_i| \bm{1}_n)) \vee \log \Phi(c_i + |A_i| \bm{1}_n)$, and $\Phi(x)$ represents the cumulative density function of the standard Gaussian distribution.
So $E \{|\log g_i(\tilde{S}|\theta^{(i)})|\} < \infty$, and (i) of Lemma \ref{passinglemma} holds.

Assumption \ref{noiseAsmpP} guarantees that the continuous differentiability of $\log g_i(\tilde{S}(\omega)|\theta^{(i)})$, and hence (ii) of Lemma \ref{passinglemma} holds. Since $\tilde{S} \in \mathcal{S}^{2n}$, 
\begin{align}\label{first}\nonumber
&\frac{\partial}{\partial a_{ij}} \log g_i(\tilde{S}|\theta^{(i)})\\\nonumber
&=\frac{\partial}{\partial a_{ij}} \log \big[(1 - \Phi(c_i - A_{i} S))^{\tilde{S}_i} \Phi(c_i - A_{i} S)^{1 - \tilde{S}_i} \big]\\
&= S_j \left( \frac{\tilde{S}_i \phi(c_i - A_i S)}{1 - \Phi(c_i - A_i S)} - \frac{(1 - \tilde{S}_i) \phi(c_i - A_i S)}{\Phi(c_i - A_iS)} \right)
\end{align}
is also bounded in $\|\theta^{(i)}\| < M$ by the assumption, where $\phi(x)$ represents the probability density function of the standard Gaussian distribution.

Similarly, 
\begin{align}\label{firstc}\nonumber
&\frac{\partial}{\partial c_i} \log g_i(\tilde{S}|\theta^{(i)})\\\nonumber
&=\frac{\partial}{\partial c_i} \log \big[(1 - \Phi(c_i - A_{i} S))^{\tilde{S}_i} \Phi(c_i - A_{i} S)^{1 - \tilde{S}_i} \big]\\
&= - \frac{\tilde{S}_i \phi(c_i - A_i S)}{1 - \Phi(c_i - A_i S)} + \frac{(1 - \tilde{S}_i) \phi(c_i - A_i S)}{\Phi(c_i - A_iS)} 
\end{align}
is bounded in $\|\theta^{(i)}\| < M$. This verifies (iii) of Lemma \ref{passinglemma}, which implies \eqref{passing1}.

Analogously, Lemma \ref{passinglemma}(i), (ii) hold for $\nabla_{\theta^{(i)}} \log g_i(\tilde{S}|\theta^{(i)})$. From \eqref{first} \eqref{firstc} we can obtain that 
\begin{align}\label{second}\nonumber
&\frac{\partial^2}{\partial a_{ij}\partial a_{ik}} \log g_i(\tilde{S}|\theta^{(i)}) \\\nonumber
&= - S_jS_k \bigg\{ \frac{\tilde{S}_i \phi(c_i-A_iS)}{(1 - \Phi(c_i-A_iS))^2} \\\nonumber
&\qquad~ \times \big[ \phi(c_i - A_i S) - (c_i - A_i S) (1 - \Phi(c_i - A_i S)) \big] \\\nonumber
&\qquad~~~ + \frac{(1 - \tilde{S}_i)\phi(c_i - A_i S)}{\Phi^2(c_i - A_iS)} \\\nonumber
&\qquad~~~~~\times \big [ (c_i - A_iS) \Phi(c_i - A_i S) + \phi(c_i - A_iS) \big]\bigg\},\\
&= - S_jS_k [ \tilde{S}_i G(A_iS - c_i) + (1 - \tilde{S}_i) G(c_i - A_i S) ],
\end{align}
\begin{align}\label{secondc}\nonumber
&\frac{\partial^2}{\partial a_{ij}\partial c_i} \log g_i(\tilde{S}|\theta^{(i)})\\
&= S_j [ \tilde{S}_i G(A_iS - c_i) + (1 - \tilde{S}_i) G(c_i - A_i S) ],
\end{align}
\begin{align}\label{secondcc}\nonumber
&\frac{\partial^2}{\partial^2 c_i} \log g_i(\tilde{S}|\theta^{(i)}) \\
&= - [ \tilde{S}_i G(A_iS - c_i) + (1 - \tilde{S}_i) G(c_i - A_i S) ],
\end{align}
where $1 \le i, j, k \le n$, $G(x) = \frac{x\phi(x)\Phi(x) + \phi^2(x)}{\Phi^2(x)}$. Lemma \ref{G} and the boundedness of $\tilde{S}$ indicate that $\tilde{S}_i G(A_iS - c_i) + (1 - \tilde{S}_i) G(c_i - A_i S)$ has the following bounds that depends only on $M$
\begin{align*}
\varepsilon_M \le \tilde{S}_i G(A_iS - c_i) + (1 - \tilde{S}_i) G(c_i - A_i S) \le \varepsilon_M',
\end{align*}
for all $1 \le i \le n$, where $0 < \varepsilon_M < \varepsilon_M'$.
So \eqref{second}, \eqref{secondc}, and \eqref{secondcc} are bounded, and (iii) of Lemma \ref{passinglemma} holds.
Consequently, \eqref{passing2} is verified.

\noindent\emph{Step 2.} Now we prove the Hessian of $E \{\log g_i(\tilde{S}|\theta^{(i)})\}$ is negative definite over $\mathbb{R}^{n + 1}$.

Step 1 indicates that \eqref{second}, \eqref{secondc}, and \eqref{secondcc} have upper bounds $-\varepsilon_M' S_jS_k$, $\varepsilon_M S_j$ and $-\varepsilon_M'$ in $\|\theta^{(i)}\| < M$. Hence, setting $\alpha^T := (S^T,-1)$, it follows that
\begin{equation*}
\begin{aligned}
\nabla^2_{\theta^{(i)}} E \{\log g_i(\tilde{S}|\theta^{(i)})\}&= E\{ \nabla^2_{\theta^{(i)}} \log g_i(\tilde{S}|\theta^{(i)})\}\\
& \le -(\varepsilon_M \wedge \varepsilon_M') E \{\alpha \alpha^T\}.
\end{aligned}
\end{equation*}
For $\bm{x} \in \mathbb{R}^{n + 1}$,
\begin{align*}
\bm{x}^T E \{\alpha \alpha^T\} \bm{x}&= E \{ \bm{x}^T \alpha \alpha^T \bm{x} \} = E \{(\alpha^T \bm{x})^2\} \ge 0.
\end{align*}
Suppose that $x_{n + 1} \not= 0$. We know from Theorem \ref{MC2prime} that $P \{S = \bm{0}_n\} > 0$. Thus, $E \{ (\alpha^T \bm{x})^2\} \ge x_{n + 1}^2 P \{S = \bm{0}_n\} > 0$. 
Now suppose that $x_{n + 1} = 0$ but $x_{i} \not= 0$ for some $1 \le i \le n$. Similarly, $P \{S = \bm{e_i}\} > 0$, and consequently $E\{ (\alpha^T \bm{x})^2\} \ge x_{i}^2 P \{S = \bm{e_i}\} > 0$. 
Therefore, the matrix $E \{\alpha \alpha^T\}$ is positive definite for fixed $\theta^{(i)}$ in $\|\theta^{(i)}\| < M$. 
Consequently, from the arbitrariness of $M$, $\nabla^2_{\theta^{(i)}} E \{\log g_i(\tilde{S} | \theta^{(i)})\}$ is negative definite over $\mathbb{R}^{n+1}$. 

\noindent\emph{Step 3.}
Note that
\begin{align*}
&\nabla_{\theta} E \bigg\{\sum_{1 \le i \le n} \log g_i(\tilde{S}|\theta^{(i)}) \bigg\}\\
&= (\nabla_{\theta^{(1)}} E \{\log g_1(\tilde{S}|\theta^{(1)})\},\dots,\nabla_{\theta^{(n)}} E \{\log g_n(\tilde{S}|\theta^{(n)})\})^T,
\end{align*}
and $\nabla^2_{\theta} E \bigg\{\sum_{1 \le i \le n} \log g_i(\tilde{S}|\theta^{(i)})\bigg\}$ is a block diagonal matrix with matrices $\nabla^2_{\theta^{(1)}} E \{\log g_1(\tilde{S}|\theta^{(1)})\}$, $\dots$, $\nabla^2_{\theta^{(n)}} E \{\log g_n(\tilde{S}|\theta^{(n)})\}$ at the diagonal line.

So $\nabla^2_{\theta} E \bigg\{\sum\nolimits_{1 \le i \le n} \log g_i(\tilde{S}|\theta^{(i)}) \bigg\}$ is negative definite over $\mathbb{R}^{{n\cdot(n+1)}}$, and $E \bigg\{\sum\nolimits_{1 \le i \le n} \log g_i(\tilde{S}|\theta^{(i)}) \bigg\}$ is strictly concave from Propositions 1.2.6 and 2.1.2 in \cite{bertsekas2003convex}.

\noindent\emph{Step 4.} Finally, we show that $\theta^*$ is a root of equation
\begin{equation*}
\nabla_{\theta} E \bigg\{\sum\nolimits_{1 \le i \le n} \log g_i(\tilde{S}|\theta^{(i)}) \bigg\} = \bm{0}_{{n\cdot(n+1)}}.
\end{equation*}
From Step $3$, it suffices to show that $(\theta^*)^{(i)} = (A^*_i~c^*_i)^T$ is a root of equation
\begin{equation*}
\nabla_{\theta^{(i)}} E \{\log g_i(\tilde{S}|\theta^{(i)}) \} = \bm{0}_{n+1},
\end{equation*}
for $1 \le i \le n$.

Compute
\begin{align*}
&E \bigg\{\frac{\partial}{\partial a_{ij}} \log g_i(\tilde{S}|\theta^{(i)}) \bigg\} = \sum_{\tilde{\bm{s}} \in \mathcal{S}^{2n}} P\{\tilde{S} = \tilde{\bm{s}}\} \frac{\partial}{\partial a_{ij}} \log g_i(\tilde{\bm{s}}|\theta^{(i)})\\
&= \sum_{\bm{s} \in \mathcal{S}^{n}, \tilde{\bm{s}}_i \in \mathcal{S}} P\{S = \bm{s}, \tilde{S}_i = \tilde{\bm{s}}_i\}\\
&\qquad~ \times \frac{\partial}{\partial a_{ij}} \log \big[(1 - \Phi(c_i - A_{i} \bm{s}))^{\tilde{\bm{s}}_i} \Phi(c_i - A_{i} \bm{s})^{1 - \tilde{\bm{s}}_i} \big]\\
&= \sum_{\bm{s} \in \mathcal{S}^{n}} \sum_{\tilde{\bm{s}}_i \in \mathcal{S}} P\{S = \bm{s}\} P\{\tilde{S}_i = \tilde{\bm{s}}_i| S = \bm{s}\}\\
&\qquad~ \times \frac{\partial}{\partial a_{ij}} \log \big[(1 - \Phi(c_i - A_{i} \bm{s}))^{\tilde{\bm{s}}_i} \Phi(c_i - A_{i} \bm{s})^{1 - \tilde{\bm{s}}_i} \big]\\
&= \sum_{\bm{s} \in \mathcal{S}^n} P\{S = \bm{s}\} \bigg[P\{D_{1, i} > c_i^* - A_i^*\bm{s}\} \cdot \frac{ \bm{s}_j \phi(c_i - A_i\bm{s})}{1 - \Phi(c_i - A_i\bm{s})}\\
&\qquad~~~~~ - P\{D_{1, i} \le c_i^* - A_i^*\bm{s}\} \cdot \frac{\bm{s}_j \phi(c_i - A_i\bm{s})}{\Phi(c_i - A_i\bm{s})} \bigg]\\
&= \sum_{\bm{s} \in \mathcal{S}^n} P\{S = \bm{s}\} \bigg[(1 - \Phi(c_i^* - A_i^*\bm{s})) \cdot \frac{ \bm{s}_j \phi(c_i - A_i\bm{s})}{1 - \Phi(c_i - A_i\bm{s})}\\
&\qquad~~~~~ - \Phi(c_i^* - A_i^*\bm{s}) \cdot \frac{\bm{s}_j \phi(c_i - A_i\bm{s})}{\Phi(c_i - A_i\bm{s})} \bigg],
\end{align*}
for $1 \le j \le n$, where the penultimate equation follows from Lemma \ref{lem1} and $\mathcal{S} = \{0, 1\}$. The above equation is zero when $\theta^{(i)} = (\theta^*)^{(i)}$. 
The argument is similar for $E \{\frac{\partial}{\partial c_i} \log g_i(\tilde{S}|\theta^{(i)})\}$. 
From step $3$, $\theta^*$ is the unique global maximum by Proposition 2.1.2 in \cite{bertsekas2003convex}. 
\hfill$\Box$

\section{Proof of Theorem \ref{mainthm2}}\label{AppendixD}

Instead of verifying the strong consistency of Algorithm \eqref{IdenAlg}, we show the consistency of the more general algorithm in Remark \ref{rmk1}, i.e., Algorithm \eqref{SAAWET},
{	which is equivalent to Algorithm \eqref{IdenAlg} by letting $M_0 = M$, where $M$ is the bound of $\|\theta_k\|$ assumed in Remark \ref{rmk1}.}
\begin{equation}\label{SAAWET}
\begin{aligned}
\theta_{k + 1} &= (\theta_k + a_k K(\theta_k, \tilde{S}_{k + 1}))\mathbb{I}_{[\|\theta_k + a_k K(\theta_k, \tilde{S}_{k + 1})\| \le M_{\sigma_k}]},\\
\sigma_k &= \sum_{i = 1}^{k - 1} \mathbb{I}_{[\|\theta_i + a_i K(\theta_i, \tilde{S}_{i + 1})\| > M_{\sigma_i}]},
\end{aligned}
\end{equation}
where $\theta_k^T = ((\theta^{(1)}_k)^T,\dots,(\theta^{(n)}_k)^T)$ is the estimate of $\theta^*$ at time step $k$, $K(\cdot,\cdot)$ is defined in \eqref{K}, $a_k$ is the step size, $\{M_k\}$ is a sequence of positive numbers increasingly diverging to $+\infty$, and $\sigma_0 = 0$.

We need the conditions below to ensure convergence.

\noindent \textbf{\emph{A1}}. $a_k > 0$, $\sum\nolimits_{k = 1}^{\infty} a_k = \infty$, $\sum\nolimits_{k = 1}^{\infty} a_k^2 < \infty$.

\noindent \textbf{\emph{A2}}. There is a continuously differentiable function (not necessarily being nonnegative) $v(\cdot) : \mathbb{R}^{{n\cdot(n+1)}} \to \mathbb{R}$ such that for $K(\theta) := E\{K(\theta, \tilde{S})\}$, where $\tilde{S}$ is subject to the stationary distribution of $\{\tilde{S}_k\}$,
\[
\underset{d_1 \le d(\theta, J) \le d_2}{\text{sup}} K^T(\theta)v_{\theta}(\theta) < 0
\]
for any $d_2 > d_1 > 0$, and $v(J) := \{v(\theta) : \theta \in J\}$ is nowhere dense where $J := \{\theta \in \mathbb{R}^{{n\cdot(n+1)}} : K(\theta) = \bm{0}_{{n\cdot(n+1)}}\}$, $d(\theta, J) = \inf_\eta \{\|\theta - \eta\| : \eta \in J\}$, and $v_{\theta}$ denotes the gradient of $v$. 
Further, $v(0) < \inf_{\|\theta\| = d_0} v(\theta)$ for some $d_0 > 0$.

\noindent \textbf{\emph{A3}}. $K(\cdot, \cdot)$ is locally Lipschitz-continuous in the first argument, i.e., for any fixed $L > 0$,
\begin{equation}\label{gz}
\|\big(K(\theta,\eta) - K(\kappa,\eta)\big) \mathbb{I}_{[\|\theta\| \le L, \|\kappa\| \le L]}\| \le c_L \| \theta - \kappa \| g(\eta),
\end{equation}
where $c_L$ is a constant depending on $L$, and $g(\eta)$ is a measurable function $\mathbb{R}^{2n} \to \mathbb{R}$.

\noindent \textbf{\emph{A4}}. 
i) $\{\tilde{S}_k\}$ is a $\phi$-mixing process, i.e., for
\[\phi_k := \sup_{n \ge 1} ~\sup_{A \in \mathcal{F}_1^n, P(A) > 0, B \in \mathcal{F}_{n+k}^{\infty}} \frac{|P(AB)-P(A)P(B)|}{P(A)},
\]
$\phi_k \to 0$ as $k \to +\infty,$ where $\mathcal{F}_i^j := \sigma(\tilde{S}_k, i \le k \le j)$.

\noindent ii)
\begin{align*}
\sup_k E \{(g^2(\tilde{S}_{k + 1}) + \|K(0, \tilde{S}_{k + 1})\|^2)|\mathcal{F}_1^k\} &:= \mu^2 < \infty,\\
E \{(g^2(\tilde{S}) + \|K(0, \tilde{S})\|^2)\} &:= \lambda^2 < \infty,
\end{align*}
and $E \mu^2 < \infty$, where $g(\cdot)$ is defined in \eqref{gz}.

\noindent iii) $\psi_k := \sup_{A \in \mathcal{B}^m} \big|P(\tilde{S}_k \in A) - P(\tilde{S} \in A)|\to 0$, $k \to \infty$.

\begin{lem}[Theorem 2.5.1 in \cite{chen2002stochastic}]\label{sa}
Assume that the above \textbf{A1}-\textbf{A4} hold. Then for $\{\theta_k\}$ generated by \eqref{SAAWET}
\[d(\theta_k, J^*) \to 0, \text{ as } k \to \infty, \text{ a.s.,}\]
where $J^*$ is a connected subset of the closure of $J$.
\end{lem}

The strong consistency of \eqref{SAAWET}, consequently \eqref{IdenAlg}, is verified by validating the conditions A1-A4 of Lemma \ref{sa} above, and we need the following lemma.

\begin{lem}[Proposition 7.8.3 in section I.7.8 of \cite{malliavin2012integration}]\label{conlemma}
Let $(\Omega, \mathscr{F}, P)$ be a probability space and $\Theta$ be a metric space, and let $f(\cdot, \cdot) : \Theta \times \Omega \to \mathbb{R}$ be a function satisfying $E |f(\theta, \omega)| < \infty$, $\forall \theta \in \Theta$. Consider $\theta_0 \in \Theta$ such that $f(\theta, \omega)$ is continuous at $\theta_0$ for almost all $\omega \in \Omega$. Assume that there exists an integrable nonnegative function $g$ and a neighborhood $\mathcal{N}$ of $\theta_0$ such that $|f(\theta, \omega)| \le g(\omega)$, $\forall (\theta, \omega) \in \mathcal{N} \times \Omega$. Then $\int_{\Omega} f(\theta, \omega) dP(\omega)$ is continuous at $\theta_0$.
\end{lem}

\noindent\emph{\textbf{Proof of Theorem \ref{mainthm2}}:} Assumption \textbf{\emph{A1}} is the same as Assumption \ref{at}. 

Let $v(\theta) = - E \{\sum_{1 \le i \le n} \log g_i(\tilde{S}|\theta^{(i)})\}$, and it is nonnegative by the definition of $g_i$ in \eqref{g_i}. 
From Step 1 in the proof of Theorem \ref{mainthm1}, $v_{\theta}(\theta) = - E \{K(\theta, \tilde{S})\} = - E \{ \nabla_{\theta} \sum_{1 \le i \le n} \log g_i(\tilde{S}|\theta^{(i)})\}$ and $E \{\|K(\theta, \tilde{S})\|\}< \infty$, $\forall \theta \in \mathbb{R}^{{n\cdot(n + 1)}}$. 
Assumption \ref{noiseAsmpP} implies that $- K(\theta, \tilde{S}(\omega))$ is continuous in $\mathbb{R}^{{n\cdot(n + 1)}}$. 
Hence, combining the local boundedness of $- K(\theta, \tilde{S})$, the continuity of $v_{\theta}(\theta)$ in $\mathbb{R}^{{n\cdot(n + 1)}}$ follows from Lemma \ref{conlemma}. 

Also, note that $J = \{K(\theta) = 0\} = \{\theta^*\}$ by Theorem \ref{mainthm1}, and
\[
\underset{d_1 \le d(\theta, J) \le d_2}{\text{sup}} K^T(\theta) v_{\theta}(\theta) = - \|E \{K(\theta, \tilde{S})\}\|^2 < 0,
\]
for any $d_2 > d_1 > 0$, because $\theta^*$ is the only root of $v(\theta)$ from Theorem \ref{mainthm1}.

From Theorem \ref{MC2prime}, $\pi_* := \min_{\tilde{s} \in \mathcal{S}^{2n}}\big\{P\{\tilde{S} = \tilde{s}\}\big\} > 0$. For the CDF of the standard Gaussian random variable, $\Phi(\cdot)$, there exist constants $M_1 < 0$ and $M_2 > 0$ such that $\Phi(x) < \exp\{-v(0) / \pi^*\}$ for $x < M_1$ and $1 - \Phi(x) < \exp\{-v(0) / \pi^*\}$ for $x > M_2$. 
Let $M = |M_1| \vee M_2$ and $d_0 = \sqrt{4n^2 + n}(M + 1)$. Then for $\|\theta_0\| = d_0$, if there exists $c_j$ such that $|c_j| \ge M + 1$, then supposing first $c_j \ge M + 1 > M$, we have that
\begin{align*}
v(\theta_0) &= - E \{ \sum_{1 \le i \le n} \log g_i(\tilde{S}|\theta^{(i)}_0)\}\\
&\ge - \sum\nolimits_{1 \le i \le n} \log g_i(\tilde{\bm{u}}|\theta^{(i)}_0) P\{\tilde{S} = \tilde{\bm{u}}\}\\
&\ge - \log g_j(\tilde{\bm{u}}|\theta^{(j)}_0) P\{\tilde{S} = \tilde{\bm{u}}\}\\
&= - \log (1 - \Phi(c_i)) P\{\tilde{S} = \tilde{\bm{u}}\} > \frac{v(0)}{\pi_*} \pi_* = v(0),
\end{align*}
where $\tilde{\bm{u}} \in \mathcal{S}^{2n}$ is a vector with $\tilde{\bm{u}}_j = 1$ and $\tilde{\bm{u}}_{n+1:2n} = \bm{0}_n$, the first and the second inequalities follow from $- \log g_i(\tilde{S}|\theta^{(i)}_0) \ge 0$, $1 \le i \le n$, and the second equation follows from the definition of $g_i$ and $\tilde{\bm{u}}$. 
If $c_j < -(M + 1)$, then choose $\tilde{\bm{u}}$ such that $\tilde{\bm{u}}_j = 0$. Hence, $v(\theta_0) \ge - \pi^* \log \Phi(c_i) > v(0)$.

If $|c_i| < M + 1$ for all $1 \le i \le n$, then there must exist $a_{ij}$ such that $|a_{ij}| \ge 2(M + 1)$. 
Otherwise, $\|\theta_0\|^2 < n^2 4(M + 1)^2 + n (M + 1)^2 = d_0^2$. 
Suppose that $\|a_{11}\| > 2(M + 1)$ for convenience, and as above suppose further that $a_{11} \ge 2(M + 1)$. Then $c_1 - a_{11} \le -(M + 1) < -M$ since $|c_i| < M + 1$. 
Thus, selecting a vector $\tilde{\bm{w}} \in \mathcal{S}^{2n}$ such that $\tilde{\bm{w}}_j = 0$ and $\tilde{\bm{w}}_{n+1:2n} = \bm{1}_n$, analogously we have that $v(\theta) \ge - \pi^* \log \Phi(c_i - a_{11}) > v(0)$. Therefore, we have showed that there exists $d_0 > 0$ such that $v(0) < \inf_{\|\theta_0\| = d_0} v(\theta_0)$ and validated \textbf{\emph{A2}}. 

Apropos of \textbf{\emph{A3}}, for $\bar{\theta}$ and $\hat{\theta}$ such that $\|\bar{\theta}\|, \|\hat{\theta}\| \le L$ with $L > 0$ fixed and $\tilde{\bm{z}} \in \mathbb{R}^{2n}$,
\begin{align}\nonumber
&\|K(\bar{\theta}, \tilde{\bm{z}}) - K(\hat{\theta}, \tilde{\bm{z}})\| \le \sum_{i = 1}^n \|K_i(\bar{\theta}^{(i)}, \tilde{\bm{z}}) - K_i(\hat{\theta}^{(i)}, \tilde{\bm{z}})\|\\\nonumber
&\le \sum_{i = 1}^n \sum_{j = 1}^{n + 1} \bigg\|\frac{\partial}{\partial \theta^{(i)}_j} \log g_i(\tilde{\bm{z}} | \theta^{(i)})\big|_{\theta^{(i)} = \bar{\theta}^{(i)}} \\\nonumber
&\qquad\qquad\qquad - \frac{\partial}{\partial \theta^{(i)}_j} \log g_i(\tilde{\bm{z}} | \theta^{(i)})\big|_{\theta^{(i)} = \hat{\theta}^{(i)}}\bigg\|\\\nonumber
&\le \sum_{i = 1}^n \sum_{j = 1}^{n + 1} \bigg\|\nabla_{\theta^{(i)}} \big(\frac{\partial}{\partial \theta^{(i)}_j} \log g_i(\tilde{\bm{z}} | \theta^{(i)})\big)\big|_{\theta^{(i)} = \tilde{\theta}^{(i)}}\bigg\|\\\nonumber
&\qquad\qquad\qquad \times \|\bar{\theta}^{(i)} - \hat{\theta}^{(i)}\|\\\nonumber
&\le \varepsilon_L h(\tilde{\bm{z}})\sum_{i = 1}^n \|\bar{\theta}^{(i)} - \hat{\theta}^{(i)}\|\\\label{g(z)}
&\le \varepsilon_L \cdot \sqrt{n} h(\tilde{\bm{z}}) \cdot \|\bar{\theta} - \hat{\theta}\|:= \varepsilon_L \cdot g(\tilde{\bm{z}}) \cdot \|\bar{\theta} - \hat{\theta}\|,
\end{align}
where the third inequality follows from the mean value theorem, $\tilde{\theta}^{(i)} = (1 - \lambda) \bar{\theta}^{(i)} + \lambda \hat{\theta}^{(i)}$ for $1 \le i \le n$ and some $\lambda \in (0, 1)$, and the fourth inequality can be obtained from the boundedness of \eqref{second}-\eqref{secondcc} in $\|\theta\| \le L$, for some bounded function $\varepsilon_L h(\tilde{\bm{z}})$, as in the proof of Theorem \ref{mainthm1}.

Since $\{\tilde{S}_k\}$ is an aperiodic irreducible finite-state Markov chain from Theorem \ref{MC2}, it is $\phi$-mixing \cite{doukhan1994mixing}. We also have that $g^2(\tilde{S}_{k + 1}) + \|K(0, \tilde{S}_{k + 1})\|^2$ and $E \{g^2(\tilde{S}) + \|K(0, \tilde{S})\|^2\}$ are bounded because $\tilde{S}_k$ takes value only in $\mathcal{S}^{2n}$. In addition, Theorem 4.9 in \cite{levin2017markov} and Theorem \ref{MC2} imply that $\psi_k \to 0$ as $k \to \infty$. Therefore, \textbf{\emph{A4}} holds, and the conclusion follows from Lemma \ref{sa} by noticing that $J = \{\theta^*\}$.
\hfill$\Box$

\section{Proof of Theorem \ref{ConvergenceRateThm}}\label{AppendixE}

Recall $K(\theta) = E \{K(\theta, \tilde{S})\}$, where $K(\theta, \tilde{S})$ is defined in \eqref{Ki}-\eqref{K} and $\tilde{S}$ is subject to the stationary distribution of $\{\tilde{S}_k\}$. We know from Theorem \ref{mainthm1} that $K(\theta)$ has a single root $\theta^*$. In addition, it is differentiable at $\theta^*$, and its Taylor expansion at $\theta^*$ is $K(\theta) = F(\theta - \theta^*) + \delta(\theta)$, where $\delta(\theta^*) = 0$ and $\delta(\theta) = o(\|\theta - \theta^*\|)$ as $\theta \to \theta^*$.

Consider the following conditions.

\noindent \textbf{\emph{A1'}}. $a_k > 0$, $a_k \to 0$ as $k \to \infty$, $\sum_{k = 1}^{\infty} a_k = \infty$, and 
\begin{equation}\label{B1}
\frac{a_k - a_{k-1}}{a_k a_{k-1}} \to \alpha \ge 0, ~k \to \infty.
\end{equation}

\noindent \textbf{\emph{A3'}}. $K(\theta)$ is measurable and locally bounded, and is differentiable at $\theta^*$ such that as $\theta \to \theta^*$
\begin{equation}\label{B4}
K(\theta) = F(\theta - \theta^*) + \delta(\theta), ~\delta(\theta^*) = 0, ~\delta(\theta) = o(\|\theta - \theta^*\|).
\end{equation}
The matrix $F$ is stable (All its eigenvalues are with negative real parts). In addition, $F + \alpha \delta I$ is also stable, where $\alpha$ and $\delta$ are given by \eqref{B1} and \eqref{B3}, respectively.

\noindent \textbf{\emph{A4'}}. For the sample path $\omega$ under consideration the observation noise $\varepsilon_k := K(\theta_{k - 1}, \tilde{S}_{k}) - K(\theta_{k-1})$ can be decomposed into two parts $\varepsilon_k = \varepsilon_k' + \varepsilon_k{''}$ such that
\begin{equation}\label{B3}
\sum_{k = 1}^{\infty} a_k^{1 - \delta} \varepsilon_k' < \infty, ~\varepsilon_k^{\prime\prime} = O(a_k^{\delta}),
\end{equation}
for some $\delta \in (0, 1]$.

\begin{lem}\label{ConvergenceRateProp}(Theorem 3.1.1 in \cite{chen2002stochastic})
Assume \textbf{A1'}, \textbf{A2}, \textbf{A3'}, and \textbf{A4'} hold. Then for those sample paths for which \eqref{B3} holds, $\theta_k$ given by \eqref{IdenAlg} converges to $\theta^*$ with the following convergence rate:
\begin{equation}\label{ConvergenceRate}
\|\theta_k - \theta^*\| = o(a_k^{\delta}),
\end{equation}
where $\delta$ is the one given in \eqref{B3}.
\end{lem}

We first {prove} a auxiliary lemma as follows.

\begin{lem}\label{poissonlemma}
For fixed $\theta \in \mathbb{R}^{{n\cdot(n+1)}}$ and $z \in \mathcal{S}^{2n}$, the series 
\begin{equation}\label{khat}
\hat{K}(\theta, \bm{z}) = \sum_{k = 0}^{\infty} \bigg(\sum_{\bm{z}' \in \mathcal{S}^{2n}} K(\theta, \bm{z}') \tilde{P}^k(\bm{z}, \bm{z}') - K(\theta) \bigg)
\end{equation}
converges, and it is a solution of the following Poisson equation
\begin{equation}\label{poisson}
K(\theta, \bm{z}) - K(\theta) = \hat{K}(\theta, \bm{z}) - \sum_{\bm{z}' \in \mathcal{S}^{2n}} \hat{K}(\theta, \bm{z}') \tilde{P}(\bm{z}, \bm{z}'),
\end{equation}
where $\tilde{P}(\cdot, \cdot)$ and $\tilde{P}^k(\cdot, \cdot)$ are the transition probability matrix and $k$-step transition probability matrix of $\{\tilde{S}_k\}$ respectively, and $\tilde{P}^0(\bm{z}, \bm{z}') = 1$ if $\bm{z}' = \bm{z}$, $\tilde{P}^0(\bm{z}, \bm{z}') = 0$ otherwise.
\end{lem}

\begin{proof}
Note that
\begin{align*}
&\bigg\| \sum_{\bm{z}' \in \mathcal{S}^{2n}} K(\theta, \bm{z}') \tilde{P}^k(\bm{z}, \bm{z}') - K(\theta) \bigg\|\\
&=\bigg\| \sum_{\bm{z}' \in \mathcal{S}^{2n}} K(\theta, \bm{z}') \tilde{P}^k(\bm{z}, \bm{z}') - K(\theta, \bm{z}') \pi(\bm{z}') \bigg\| \\
&\le \sum_{\bm{z}' \in \mathcal{S}^{2n}} \|K(\theta, \bm{z}')\| \cdot |\tilde{P}^k(\bm{z}, \bm{z}') - \pi(\bm{z}')| \\
&\le \max_{\bm{z}' \in \mathcal{S}^{2n}} \|K(\theta, \bm{z}')\| \cdot \sum_{\bm{z}' \in \mathcal{S}^{2n}} |\tilde{P}^k(\bm{z}, \bm{z}') - \pi(\bm{z}')| \\
&\le \max_{\bm{z}' \in \mathcal{S}^{2n}} \|K(\theta, \bm{z}')\| \cdot C_1 \rho^k,
\end{align*}
where $\pi$ is the stationary distribution of $\{\tilde{S}_k\}$, and the last inequality follows from the convergence theorem of finite-state Markov chains (Theorem 4.9 in \cite{levin2017markov}) for some $C_1 > 0$, $\rho \in (0, 1)$, and any $\bm{z} \in \mathcal{S}^{2n}$. Hence, 
\begin{align*}
\|\hat{K}(\theta, \bm{z})\| &\le \max_{\bm{z}' \in \mathcal{S}^{2n}} \|K(\theta, \bm{z}')\| \cdot \sum_{k = 0}^{\infty} C_1 \rho^k \\
&:= C_2 \cdot \max_{\bm{z}' \in \mathcal{S}^{2n}} \|K(\theta, \bm{z}')\|,
\end{align*}
where $C_2$ is a positive constant not relying on $\theta$.

The equation \eqref{poisson} is obtained by noticing that
\begin{align*}
&\sum_{\bm{z}' \in \mathcal{S}^{2n}} \hat{K}(\theta, \bm{z}') \tilde{P}(\bm{z}, \bm{z}')\\
&= \sum_{\bm{z}' \in \mathcal{S}^{2n}} \sum_{k = 0}^{\infty} \sum_{\bm{z}'' \in \mathcal{S}^{2n}} (K(\theta, \bm{z}'') \tilde{P}^k(\bm{z}', \bm{z}'') - K(\theta)) \tilde{P}(\bm{z}, \bm{z}') \\
&= \sum_{k = 0}^{\infty} \sum_{\bm{z}'' \in \mathcal{S}^{2n}} \sum_{\bm{z}' \in \mathcal{S}^{2n}} (K(\theta, \bm{z}'') \tilde{P}^k(\bm{z}', \bm{z}'') - K(\theta)) \tilde{P}(\bm{z}, \bm{z}')\\
&= \sum_{k = 0}^{\infty} \bigg(\sum_{\bm{z}'' \in \mathcal{S}^{2n}} K(\theta, \bm{z}'') \tilde{P}^{k+1}(\bm{z}, \bm{z}'') - K(\theta) \bigg)\\
&= \hat{K}(\theta, \bm{z}) - (K(\theta, \bm{z}) - K(\theta)).
\end{align*}
\end{proof}

\noindent\emph{\textbf{Proof of Theorem \ref{ConvergenceRateThm}}:} First note that $\frac{a_k - a_{k-1}}{a_k a_{k-1}} \to 0$ when $\beta \in (0, 1/3)$. Thus \textbf{\emph{A1'}} holds with $\alpha = 0$. In addition, $K(\theta)$ is the same for \textbf{\emph{A2}} as in Appendix \ref{AppendixD}, which has been verified in the proof of Theorem \ref{mainthm2}. From the definition of $K(\theta)$, we know that $F$ in \eqref{B4} is in fact the Hessian of $E \{\sum_{1 \le i \le n} \log g_i(\tilde{S}|\theta^{(i)})\}$ at $\theta^*$. It follows that $F$ is negative definite and consequently stable from the proof of Theorem \ref{mainthm1}. Thus \textbf{\emph{A3'}} holds as $\alpha = 0$ for $\beta \in (0, 1/3)$.

Now we show that \textbf{\emph{A4'}} holds a.s. for $\delta \in (0, 1/2)$ by using \eqref{poisson} and decomposing the noise into three parts:
\begin{align*}
\varepsilon_{k} &= K(\theta_{k-1}, \tilde{S}_{k}) - K(\theta_{k-1}) \\
&= \hat{K}(\theta_{k-1}, \tilde{S}_{k}) - \sum\nolimits_{\bm{z} \in \mathcal{S}^{2n}} \hat{K}(\theta_{k-1}, \bm{z}) \tilde{P}(\tilde{S}_{k}, \bm{z})\\ 
&= I_{k}^{(1)} + I^{(2)}_{k} + I^{(3)}_{k},
\end{align*}
where for $k \ge 2$,
\begin{align*}
I_{k}^{(1)} &= \hat{K}(\theta_{k-1}, \tilde{S}_{k}) - \sum\nolimits_{\bm{z} \in \mathcal{S}^{2n}} \hat{K}(\theta_{k-1}, \bm{z}) \tilde{P}(\tilde{S}_{k-1}, \bm{z}),\\
I_{k}^{(2)} &= \sum\nolimits_{\bm{z} \in \mathcal{S}^{2n}} \hat{K}(\theta_{k-1}, \bm{z}) \tilde{P}(\tilde{S}_{k-1}, \bm{z}) \\
&\qquad~- \sum\nolimits_{\bm{z} \in \mathcal{S}^{2n}} \hat{K}(\theta_{k - 2}, \bm{z}) \tilde{P}(\tilde{S}_{k-1}, \bm{z}),\\
I_{k}^{(3)} &= \sum\nolimits_{\bm{z} \in \mathcal{S}^{2n}} \hat{K}(\theta_{k - 2}, \bm{z}) \tilde{P}(\tilde{S}_{k-1}, \bm{z}) \\
&\qquad~- \sum\nolimits_{\bm{z} \in \mathcal{S}^{2n}} \hat{K}(\theta_{k-1}, \bm{z}) \tilde{P}(\tilde{S}_{k}, \bm{z}).
\end{align*}

It follows that $\hat{K}(\theta_{k}, \tilde{S}_{k+1})I_{[\|\theta_k\| \le N]}$ is bounded a.s. for fixed $N > 0$ from Lemma \ref{poissonlemma}. Denote $\mathcal{F}_i^j := \sigma\{\tilde{S}_k, i \le k \le j\}$, it holds that for $N > 0$
\begin{align*}
&E\{\hat{K}(\theta_{k}, \tilde{S}_{k+1})I_{[\|\theta_k\| \le N]}|\mathcal{F}_1^k\}(\omega)\\
&=\int \hat{K}(\theta_k(\omega), \bm{z})I_{[\|\theta_k\| \le N]} dF_{k+1}^{\omega}(\bm{z}; \mathcal{F}_1^k)\\
&=\int \hat{K}(\theta_k(\omega), \bm{z})I_{[\|\theta_k\| \le N]} dF_{k+1}^{\omega}(\bm{z}; \mathcal{F}_k^k)\\
&= \sum\nolimits_{\bm{z} \in \mathcal{S}^{2n}} \hat{K}(\theta_{k}(\omega), \bm{z})I_{[\|\theta_k\| \le N]} \tilde{P}(\tilde{S}_{k}(\omega), \bm{z}),
\end{align*}
where $F_{k+1}^{\omega}(\cdot; \mathcal{F}_1^k)$ is the conditional distribution of $\tilde{S}_{k+1}$ given $\mathcal{F}_1^k$, and the second equality follows from the Markov property of $\{\tilde{S}_k\}$. Thus 
\begin{align*}
I^{(1, N)}_{k} &: =\hat{K}(\theta_{k-1}, \tilde{S}_{k})I_{[\|\theta_{k-1}\| \le N]} \\
&\quad~~~- \sum\nolimits_{\bm{z} \in \mathcal{S}^{2n}} \hat{K}(\theta_{k-1}, \bm{z})I_{[\|\theta_{k-1}\| \le N]} \tilde{P}(\tilde{S}_{k-1}, \bm{z})
\end{align*}
is a martingale difference sequence for any $N>0$. For $\delta \in (0, 1 - \frac1{2(1 - \beta)})$ and $\beta \in (0, 1/3)$, 
\[2 (1 - \beta)(1 - \delta) \in (1, 2(1 - \beta)),\]
so $\sum_{k=1}^{\infty} a_k^{2(1 - \delta)} < \infty$ for $a_k = a/(k^{1-\beta} + \gamma)$, and $\sum_{k=1}^{\infty} a_k^{1-\delta} I^{(1,N)}_{k} < \infty$ for $N > 0$ by Theorem B.6.1 in \cite{chen2002stochastic}.

From Theorem \ref{mainthm2}, for a fixed sample path $\omega \in \Omega_0$ with $P(\Omega_0) = 1$, $\theta_k(\omega) \to \theta^*$ as $k \to \infty$. So there exists an integer $L = L(\omega) > 0$ such that for all $k \ge 0$, $\|\theta_k(\omega)\| \le L$. Hence set $N = L$, we have that for $\delta \in (0, 1 - \frac1{2(1 - \beta)})$
\[
\sum_{k=1}^{\infty} a_k^{1-\delta} I^{(1)}_{k}(\omega) = \sum_{k=1}^{\infty} a_k^{1-\delta} I^{(1,L)}_{k}(\omega) < \infty.
\]
To analyze $I_{k}^{(2)}$, first we have for $\bar{\theta}$, $\hat{\theta}$ with $\|\bar{\theta}\|, \|\hat{\theta}\| \le L$
\begin{align*}
&\|\hat{K}(\bar{\theta}, \bm{z}) - \hat{K}(\hat{\theta}, \bm{z})\|\\
&= \bigg\|\sum_{k = 0}^{\infty} \sum_{\bm{z}' \in \mathcal{S}^{2n}} (K(\bar{\theta}, \bm{z}') - K(\hat{\theta}, \bm{z}'))(\tilde{P}^k(\bm{z}, \bm{z}') - \pi(\bm{z}'))\bigg\|\\
&\le \sum_{k = 0}^{\infty} \sum_{\bm{z}' \in \mathcal{S}^{2n}} \|K(\bar{\theta}, \bm{z}') - K(\hat{\theta}, \bm{z}')\| \cdot |\tilde{P}^k(\bm{z}, \bm{z}') - \pi(\bm{z}')|\\
&\le \sum_{k = 0}^{\infty} \sum_{\bm{z}' \in \mathcal{S}^{2n}} \|\bar{\theta} - \hat{\theta}\| \varepsilon_L \cdot g(\bm{z}') \cdot |\tilde{P}^k(\bm{z}, \bm{z}') - \pi(\bm{z}')|\\
&= \varepsilon_L \cdot \max_{\bm{z} \in \mathcal{S}^{2n}}  g(\bm{z}) \cdot \|\bar{\theta} - \hat{\theta}\| \cdot \sum_{k = 0}^{\infty} \sum_{\bm{z}' \in \mathcal{S}^{2n}} |\tilde{P}^k(\bm{z}, \bm{z}') - \pi(\bm{z}')|\\
&\le \varepsilon_L C_2 \max_{\bm{z} \in \mathcal{S}^{2n}} g(\bm{z}) \cdot \|\bar{\theta} - \hat{\theta}\|,
\end{align*}
where the second inequality follows from \eqref{g(z)} in the proof of \textbf{\emph{A3}} of Theorem \ref{mainthm2}, the last inequality is obtained as in Lemma \ref{poissonlemma} with the constant $C_2$, and $\pi$ is the stationary distribution of $\{\tilde{S}_k\}$.

Hence, for the fixed sample path $\omega$ such that $\|\theta_k(\omega)\| \le L$, $\forall k \ge 0$,
\begin{align*}
&\bigg\|\sum_{\bm{z} \in \mathcal{S}^{2n}} \hat{K}(\theta_{k}, \bm{z}) \tilde{P}(\tilde{S}_{k}, \bm{z}) - \sum\nolimits_{\bm{z} \in \mathcal{S}^{2n}} \hat{K}(\theta_{k - 1}, \bm{z}) \tilde{P}(\tilde{S}_{k}, \bm{z})\bigg\|\\
&\le 
\sum_{\bm{z} \in \mathcal{S}^{2n}} \|\hat{K}(\theta_{k}, \bm{z})I_{[\|\theta_{k}\| \le L]} - \hat{K}(\theta_{k - 1}, \bm{z})I_{[\|\theta_{k-1}\| \le L]}\| \\
&\qquad\qquad~ \times \tilde{P}(\tilde{S}_{k}, \bm{z})\\
&\le C_2 \cdot \|\theta_{k} - \theta_{k-1}\| \cdot \max_{\bm{z} \in \mathcal{S}^{2n}} g(\bm{z}) \cdot \sum_{\bm{z} \in \mathcal{S}^{2n}} \tilde{P}(\tilde{S}_{k}, \bm{z})\\
&= C_2 a_k \|K(\theta_{k-1}, \tilde{S}_k)\| \cdot \max_{\bm{z} \in \mathcal{S}^{2n}} g(\bm{z})\\
&= C_2 a_k \|K(\theta_{k-1}, \tilde{S}_k)I_{[\|\theta_{k-1}\| \le L]}\| \cdot \max_{\bm{z} \in \mathcal{S}^{2n}} g(\bm{z}) \le C_3 a_k,
\end{align*}
where the second inequality follows from the above assertion, and $C_3$ is a constant.

So noticing that $(1 - \beta)(2 - \delta) > 1$ for $\beta \in (0, 1/3)$ and $\delta \in (0, 1/2)$, we have that $\sum_{k = 1}^{\infty} a_k^{1 - \delta} I_{k}^{(2)}(\omega) \le C_3 \sum_{k = 1}^{\infty} a_k^{2 - \delta} < \infty$ for $\delta \in (0, 1/2)$ and $a_k = a/(k^{1-\beta} + \gamma)$.

As for $I_{k}^{(3)}$, rewrite $\sum_{k = 1}^{\infty} a_k^{1 - \delta} I_{k}^{(3)}(\omega)$ for the fixed sample path $\omega$ such that $\|\theta_k(\omega)\| < L$, $\forall k > 0$, as 
\begin{align*}
& \bigg\|\sum_{k = 2}^{\infty} a_k^{1 - \delta} I_{k}^{(3)} \bigg\| \\
&= \bigg\|\sum_{k = 0}^{\infty} (a_{k+2}^{1 - \delta} - a_{k+1}^{1-\delta}) \bigg(\sum\nolimits_{\bm{z} \in \mathcal{S}^{2n}} \hat{K}(\theta_{k}, \bm{z}) \tilde{P}(\tilde{S}_{k+1}, \bm{z}) \bigg)\bigg\|\\
&\le \sum_{k = 0}^{\infty} |a_{k+2}^{1 - \delta} - a_{k+1}^{1-\delta}| \bigg(\sum\nolimits_{\bm{z} \in \mathcal{S}^{2n}} \|\hat{K}(\theta_{k}, \bm{z})\| \tilde{P}(\tilde{S}_{k+1}, \bm{z}) \bigg)\\
&= \sum_{k = 0}^{\infty} |a_{k+2}^{1 - \delta} - a_{k+1}^{1-\delta}|\\
&\qquad~~~ \times \bigg(\sum\nolimits_{\bm{z} \in \mathcal{S}^{2n}} \|\hat{K}(\theta_{k}, \bm{z})I_{[\|\theta_{k}\| \le L]}\| \tilde{P}(\tilde{S}_{k+1}, \bm{z}) \bigg)\\
&\le C_4 \sum_{k = 1}^{\infty} |a_{k}^{1 - \delta} - a_{k+1}^{1-\delta}| = O(\sum_{k = 1}^{\infty} \frac1{k^{1 + (1 - \beta)(1 - \delta)}}),
\end{align*}
where the second inequality follows from the proof of Lemma \ref{poissonlemma} for a constant $C_4$, and the last equation is obtained from the fact that for $\delta \in (0, 1/2)$, $a_k = \frac{a}{k^{1 - \beta} + \gamma}$ and $\beta \in (0, 1/3)$
\begin{align*}
&a_{k}^{1 - \delta} - a_{k+1}^{1-\delta} \\
&= a^{1-\delta} \frac{1}{((k+1)^{1 - \beta} + \gamma)^{1-\delta}} \bigg(\big(\frac{(k+1)^{1 - \beta} + \gamma}{k^{1 - \beta} + \gamma}\big)^{1 - \delta} - 1\bigg)\\
&\sim a^{1-\delta} \frac{1}{k^{(1 - \beta)(1-\delta)}} \bigg(\big(1 + \frac1k)^{(1 - \beta)(1 - \delta)} - 1\bigg)\\
&= O(\frac1{k^{1 + (1 - \beta)(1-\delta)}}),
\end{align*}
where for two sequences $\{\alpha_k\}$ and $\{\beta_k\}$ with $\beta_k \not= 0$, $k \ge 1$, $\alpha_k \sim \beta_k$ means that $\lim_{k \to \infty} \alpha_k/\beta_k = 1$.

To sum up, we have shown that $\sum_{k = 1}^{\infty} a_k^{1 - \delta} \varepsilon_k(\omega) < \infty$ for $\delta \in (0, 1 - \frac1{2(1 - \beta)})$ and $\beta \in (0, 1/3)$. By Lemma \ref{ConvergenceRateProp}, $\|\theta_k(\omega) - \theta^*\| = o(a_k^{\delta}) = O(k^{-\eta})$, $\eta = (1 - \beta)\delta$. The conclusion follows from $\eta = (1 - \beta)\delta \in (0, \frac12 - \beta)$ for $\delta \in (0, 1 - \frac1{2(1 - \beta)})$.

When $\beta = 0$, $\frac{a_k - a_{k-1}}{a_ka_{k-1}} \to \frac1a = \alpha$. Similar to the above argument, we know that \textbf{\emph{A4'}} holds a.s. for $\delta \in (0, 1/2)$. According to \textbf{\emph{A3'}}, $F + \alpha \delta I$ has to be stable. But the maximum eigenvalue of $F$ depends on the parameter vector $\theta^*$. Nevertheless, from the negative definiteness of $F$, there exists $\delta' \in (0, 1/2)$ such that $F + \alpha \delta' I < 0$ for fixed $F$. So for $\delta \in (0, \delta')$ we have that $\|\theta_k - \theta^*\| = o(k^{-\delta})$.
\hfill$\Box$\\

\ifCLASSOPTIONcaptionsoff
\newpage
\fi

\addtolength{\textheight}{-7.5cm}

\bibliographystyle{ieeetr}
\bibliography{interpersonal}

\end{document}